\documentclass[pra,twocolumn,preprintnumbers,amsmath,amssymb,floatfix,superscriptaddress,noshowpacs]{revtex4}

\usepackage[english]{babel}

\usepackage[mathletters]{ucs}
\usepackage[utf8x]{inputenc}

\usepackage{amsmath}
\usepackage{amssymb}
\usepackage{bbm}
\usepackage{graphicx}
\usepackage{graphics}

\usepackage{psfrag}
\usepackage{hyperref}
\usepackage{multirow}
\usepackage{xcolor}
\usepackage[sort&compress]{natbib}

\hypersetup{
  colorlinks,
  linkcolor={red!75!black},
  citecolor={blue!75!black},
  urlcolor={blue!75!black},
  pdftitle={Dimensional crossover in ultracold Fermi gases from Functional Renormalisation},
  pdfauthor={Faigle-Cedzich, Pawlowski, Wetterich},
  pdfkeywords={RG} {Dimensional crossover} {BCS-BEC crossover},
  bookmarksopen=true,
  bookmarksopenlevel=2,
  bookmarksnumbered=true
}

\newcommand{\sgn}{\mbox{sgn}}

\def\0#1#2{\frac{#1}{#2}}

\newcommand{\Eq}[1]{Eq.~(\ref{#1})}
\def\eq#1{(\ref{#1})}
\def\Eq#1{Eq.~(\ref{#1})}
\def\CV{{\mathcal V}}

\newcommand{\Tr}{\mathrm{Tr}}

\newcommand{\iu}{\mathrm{i}\mkern1mu}
\newcommand{\tr}{\text{Tr}}
\newcommand{\detB}{\det_{\text{B}}^Q}
\newcommand{\detF}{\det_{\text{F}}^Q}

\newcommand{\detFsim}{\det_{\text{F}}}
\newcommand{\nfo}{N_1^{(F)}}
\newcommand{\nft}{N_2^{(F)}}
\newcommand{\nftt}{N_3^{(F)}}

\newcommand{\rFtwo}{R_ψ^{(2)}}
\newcommand{\rBone}{R_ϕ^{(1)}}
\newcommand{\rBtwo}{R_ϕ^{(2)}}

\usepackage{amsfonts}
\newcommand{\overbar}[1]{\mkern 1.5mu\overline{\mkern-1.5mu#1\mkern-1.5mu}\mkern1.5mu}

\begin{document}

\hypersetup{pdftitle={Dimensional crossover in ultracold Fermi gases from Functional Renormalisation}}
\title{Dimensional crossover in ultracold Fermi gases from Functional Renormalisation}
\date{\today}
\author{Bruno M. Faigle-Cedzich}
\affiliation{Institute for Theoretical Physics, Heidelberg University, D-69120 Heidelberg, Germany}
\author{Jan M. Pawlowski}
\affiliation{Institute for Theoretical Physics, Heidelberg University, D-69120 Heidelberg, Germany}
\affiliation{ExtreMe Matter Institute EMMI, GSI Helmholtzzentrum für Schwerionenforschung mbH, D-64291 Darmstadt, Germany}
\author{Christof Wetterich}
\affiliation{Institute for Theoretical Physics, Heidelberg University, D-69120 Heidelberg, Germany}

\begin{abstract}
  We investigate the dimensional crossover from three to two
  dimensions in an ultracold Fermi gas across the whole BCS-BEC
  crossover. Of particular interest is the strongly interacting regime
  as strong correlations and pair fluctuations are more 
  pronounced in reduced dimensions. Our results
  are obtained from first principles within the framework of the
  functional renormalisation group (FRG). Here, the confinement of the
  transverse direction is imposed by means of periodic boundary
  conditions. We calculate the equation of state, the gap parameter at
  zero temperature and the superfluid transition temperature across a
  wide range of transversal confinement length scales. Particular
  emphasis is put on the determination of the finite temperature phase
  diagram for different confinement length scales. In the end, our
  results are compared with recent experimental observations and we discuss them in
  the context of other theoretical works.

\end{abstract}


\maketitle


\section{Introduction}

Lower-dimensional systems are of particular interest both in condensed
matter and statistical physics as they feature a pronounced influence
of fluctuations. Furthermore, they are of experimental and
technological importance with examples ranging from high temperature
superconductors over layered semiconductors to graphene.  To
disentangle the effects of the dimensionality from other many-body
physics effects constitutes a key challenge in the study of systems of
reduced dimensionality.

With the recent progress in trapping ultracold atomic gases in
quasi-two-dimensional geometries \cite{2007AdPhy..56..243L,Bloch:2008zzb} both zero
\cite{2016PhRvL.116d5303B,2018PhRvL.121l0401H,2018PhRvA..97f3622T} as
well as finite temperature effects
\cite{PhysRevLett.105.230408,2016PhRvL.116d5303B,PhysRevLett.114.230401,2016PhRvL.116d5302F,
2011PhRvA..84f1606P,Murthy_2018,2018PhRvA..97e3612M}
have been measured.  Hereby, strongly anisotropic trapping potentials on
the one hand and one-dimensional optical lattices one the other hand
allow for the experimental realisation of quasi-two-dimensional
quantum gases.

For example, the algebraic correlations associated with the Berezinskii–Kosterlitz–Thouless (BKT) phase transition in
\mbox{(quasi-)} two-dimensional systems have been observed in bosonic \cite{hadzibabic:2006,
2009PhRvL.102q0401C,PhysRevLett.105.230408,
2011PhRvA..84f1606P,2012NatPh...8..645D,2015PhRvL.114y5302F}, as well
as fermionic systems \cite{PhysRevLett.114.230401,2015PhRvL.115a0401M}.
In addition, \mbox{(quasi-)} two-dimensional systems exhibit the breaking
of the scale invariance in the strongly interacting regime of the
BCS-BEC crossover. Here, extensive progress both in theory
\cite{1997PhRvA..55..853P,2010PhRvL.105i5302O,2012PhRvL.109m5301T,Hofmann:2012np,
2012PhRvA..86d3609G,2013PhRvA..88d3636C,2018PhRvA..97c3630D,2019PhRvL.122g0401H},
as well as in experiment \cite{2018PhRvL.121l0401H,2018PhRvL.121l0402P,2019Sci...365..268M} has been achieved in recent years.

Due to an insufficient degree of anisotropy in the experimental setup
one may not be restricted to a particular dimension, but finds oneself
in a dimensional crossover without a well-defined dimensionality.
Apart from being an undesired effect for the investigation of pure
two-dimensional systems, the crossover may also lead to new materials
with physically interesting properties.

We concentrate here on ultracold Fermi gases. A comparable
quasi-two-dimensional setup has been studied in \cite{PhysRevA.88.023612,PhysRevB.90.214503}
in a mean-field approach, for a Fermi gas at unitarity and zero
temperature in \cite{PhysRevA.84.063616}, using the Luttinger-Ward approach 
in two dimensions in \cite{2014PhRvL.112m5302B} and using Quantum Monte Carlo (QMC) calculations in
two dimensions in \cite{PhysRevA.92.033603}.
Furthermore, two-dimensional fermionic systems have been addressed in
\cite{Marsiglio_2015,Wu_2015,He_2015,2016PhRvB..93a4519B,Salasnich_2016,Mulkerin_2017,2018IJMPB..3240022B,Wu_2020}.
Fermi gases, typically a
system of ${}^6\text{Li}$ or ${}^{40}\text{K}$, constitute a rich
physical system as their interatomic interactions may be altered via a
Feshbach resonance. For a large negative value of the three-dimensional
scattering length $a_{\text{3D}}$ the fermions form large
spatially overlapping Cooper pairs below a critical temperature (BCS
limit). On the other hand, for large positive scattering lengths, the
fermions form tightly bound molecular dimers which condense into a
Bose-Einstein condensate (BEC) at sufficiently low temperatures.

Apart from featuring the transition to the superfluid phase   
the normal-state 'pseudogap' behaviour can also be studied within the BCS-BEC crossover.
Here, the onset of superfluidity and pairing occurs at different 
temperatures, i.e. the density of states is partially gapped and the dispersion
relation is BCS-like for a range of temperatures above the critical temperature.
The system essentially retains some features of the broken 
superfluid phase also in the symmetric normal phase without exhibiting superfluidity. 
This pairing at high temperatures has been studied both experimentally, 
e.g. in \cite{Gaebler_2010,Murthy_2018}, as well as theoretically,
e.g. in \cite{Tsuchiya_2009,Matsumoto_2018,Richie-Halford:2020ooz}.

Moreover, a BCS-BEC crossover can also be found in confined superconducting
systems, where the crossover is induced by tuning the chemical potential to 
a band edge in multi-band superconductors. The size-induced molecule-like pairing 
has both been studied theoretically \cite{Chen_2012,Guidini_2014,Guidini_2016,Tajima_2020}, 
as well as experimentally \cite{Rinotte_2017}.
Here, the confinement of superconducting materials (e.g., in the form of monolayer systems)
result in shape resonances where an increased temperature, gap, as well as intrapair 
correlation length are present. The step in the density of states gives
rise to a change in the topology of the Fermi surface, a so-called Lifshitz transition, 
and is another factor in an increased critical transition temperature \cite{Blatt_1963,Thompson_1963,Innocenti_2010,Bianconi_2014,Pinto_2018,Eom_2006}.
For a 1d-2d crossover see e.g. \cite{Perali_1996}.

The BCS-BEC crossover has been studied extensively in three dimensions
using functional renormalisation group (FRG) techniques
\cite{Diehl:2005an, Diehl:2005ae, Diehl:2007ri, Diehl:2007th,
  Floerchinger:2008qc,Floerchinger:2009fp,Floerchinger:2009pg,Diehl:2009ma,
  Boettcher:2012cm,Boettcher:2012dh,Schnoerr:2013bk,PhysRevA.89.053630,
  Boettcher:2014tfa,Roscher:2015xha,2015PhLB..742...86B}.
 Finite size effects have been investigated in both cold atom systems 
 \cite{Ku:2008vk,PhysRevA.84.063616} and
 in quantum chromodynamics (QCD) \cite{Braun:2011iz,Braun:2012zz,Tripolt:2013zfa}.

In this work we study the dimensional crossover from three to two
spatial dimensions for ultracold Fermi gases by means of the functional
renormalisation group, for a study of non-relativistic bosonic systems
see \cite{2016PhRvA..93f3631L}. In particular, we are interested in
the critical temperature for the superfluid transition over the
BCS-BEC-crossover in dependence of the dimensionality.

The dimensional crossover is achieved by compactifying the
‘transverse’ $z$-direction by a potential well of length $L$.
We discuss (anti-)periodic boundary conditions,
as well as a confinement to a box with boundaries fixed to zero.
The compactification leads to a discrete momentum
spectrum in $z$-direction. 
The choice of the boundary conditions is crucial for a well-defined
two-dimensional limit. It also influences the 
mapping between three- and two-dimensional parameters of the Fermi
gas. Both aspects are discussed in detail in Section \ref{sec:BoundaryConditions}.

This paper is organised as follows: In Section \ref{sec:model} we introduce
the ultracold Fermi gas and the functional renormalisation group (FRG) in the dimensional
crossover. In particular, we discuss the aspect of boundary conditions for 
a dimensional reduction. The truncation used within the FRG, as well as the
initial conditions are presented in Section \ref{sec:runningCouplings}. In
Section \ref{sec:zeroT} the results for the equation of state and
the gap parameter in the dimensional crossover at zero temperature are discussed. 
The finite temperature phase diagrams with respect to the dimensionality are 
addressed in Section \ref{sec:dimXover}. We conclude in Section \ref{sec:conclusions}.
Some technical details are deferred to Appendix \ref{ap:IR-dep}-\ref{ap:floweq}.

\section{Model and Functional Renormalisation}\label{sec:model}

\subsection{Model}

Close to a broad Feshbach resonance, as found in quantum gases consisting of ${}^6\text{Li}$ and ${}^{40}\text{K}$, details of the atomic interactions in ultracold Fermi gases become irrelevant for the description of the macrophysics. The system can then be described by a universal action
\begin{align}\label{eq:action_micro}
\begin{split}
	S[ψ]=\int_X\left[\sum_{σ=1,2}\! ψ_σ^{*}(\partial_τ-\nabla^2-\overbar{μ})ψ_σ+\overbar{λ}_{ψ} ψ^*_1 ψ^*_2 ψ_2 ψ_1 \right]\,,
\end{split}
\end{align}
where $ψ_σ$ and $ψ^*_σ$ denote Grassmann fermions in the hyperfine state $σ=1,2$. We introduce $X=(τ,\vec{x})$ with $τ$ being the Euclidean time and $\int_X=\int_0^β dτ \int d^{d}x$ with spatial dimension $d$. Moreover, the chemical potential $\overbar{μ}$ and the four-Fermi coupling $\overbar{λ}_ψ \rightarrow λ_ψ=8\,π\,a_{\text{3D}}$ are related to the physical chemical potential and the scattering length through an appropriate vacuum renormalisation.

We use $\hbar=k_B=2M=1$ with $M$ being the mass of the fermionic atoms.
For sufficiently low temperatures, the ultracold Fermi gas may develop many-body instabilities resulting in the formation of a macroscopic anomalous self-energy $Δ$ which is related to the non-vanishing expectation value $\left<ψ_1\,ψ_2\right>$. This is signaled by a divergence of the frequency- and momentum-dependent four-Fermi vertex at lower momentum scales and causes the breaking of the global $U(1)$-symmetry.
In particular, in the strongly coupled regime, i.e. for a diverging three-dimensional s-wave scattering length $a_{\text{3D}}$, the quantitative determination of this phase transition is complicated by the frequency and momentum dependence of the vertex. In order to resolve this difficulty, a scale-dependent treatment in the path integral formulation is appropriate.

The starting point is the grand canonical partition function of the system
\begin{align}
	Z[η,\overbar{η}]=\int \mathcal{D}\overbar{ψ}\,\mathcal{D}ψ\,e^{-S[ψ,\overbar{ψ}]-\overbar{η}\cdotψ+\overbar{ψ}\cdot η}\,.
\end{align}

In order to exclude redundancies included in the grand canonical partition function, the effective action may be introduced as the Legendre transform of the Wigner functional $W[η,\overbar{η}]=\log{Z[η,\overbar{η}]}$
\begin{align}
	Γ[ψ,\overbar{ψ}]=\int_X \left(\overbar{ψ}_X\,η_X+\overbar{η}_X\,ψ_X\right)-W[η,\overbar{η}]\,.
\end{align}

\subsection{Functional renormalisation}

The non-perturbative functional renormalisation group (FRG) allows for a scale-dependent study of physical systems and theoretical models.
It is a modern implementation of Wilson's RG and enables one to go beyond perturbative methods, i.e. it is also applicable in strongly-correlated regimes. The FRG is based upon an exact functional flow-equation of a coarse-grained effective action (or Gibb's free energy in the language of statistical physics) which allows for including (thermal and quantum) fluctuations on all scales. It encompasses both Bogoliubov theory and the hydrodynamic approach of Popov and is inherently free of divergences \cite{Dupuis:2020fhh}.
Functional renormalisation proceeds in the same spirit as other functional methods used for the problem of dimensional crossover, e.g. in \cite{2014PhRvL.112m5302B,Wu_2015,2016PhRvB..93a4519B,He_2015,2018IJMPB..3240022B}. It has the advantage that several effects can be included simultaneously, and all known limits are directly realised.\\

For the scale-dependent treatment the integration is grouped in frequency and momentum shells according to
\begin{align}
	q_0^2+\left(\vec{q}\,^2-μ\right)^2 \simeq k^4
\end{align}
with external momentum scale $k$. The full grand canonical partition function is obtained by successively integrating over the corresponding frequency and momentum shells starting at $k=\infty$ and arriving in the end at $k=0$.

The microscopic action in \eqref{eq:action_micro} is related to an ultraviolet (UV) momentum scale $k=Λ$ at length scales much smaller than the van der Waals length $\ell_{\text{vdW}}$. However, the relevant physics takes place at scales $\ll Λ$ where the thermal and quantum fluctuations are included.
To incorporate these fluctuations and furthermore to obtain results in the strongly coupled regime the above scale-dependent procedure is implemented via the functional renormalisation group (FRG), which includes these fluctuations successively at each momentum scale $k$. Introducing the scale-dependent partition function
\begin{align}
	Z_k[η,\overbar{η}]=\int\mathcal{D}\overbar{ψ}\,\mathcal{D}ψ\,e^{-S[ψ]-ΔS[ψ]-\overbar{η}\cdotψ+\overbar{ψ}\cdot η}
\end{align}
the suppression of the low momentum fluctuations ${ω,\vec{q}\ll k^2}$ is incorporated via a mass-like infrared modification of the dispersion relation. In practice, a regulator or cutoff term $ΔS_k[ψ]$ is added to the microscopic action $S[ψ]$ being quadratic in the fields
\begin{align}
	ΔS[ψ]=\int_Q \sum_{σ=1,2} ψ_σ(-Q)\,R_ψ(Q)\,ψ_σ(Q)\,.
\end{align}
The regulator $R_k(Q)$ may be chosen freely with the requirements
\begin{align}
	\lim_{q^2/k^2\rightarrow 0}\,R_k(Q)=k^2\,,\quad \lim_{q^2/k^2\rightarrow \infty}\,R_k(Q)=0\,.
\end{align}
The scale-dependent effective action $Γ_k$ can be defined accordingly.
Starting at $Γ_Λ=S$, the full effective action is reached after the inclusion of all fluctuations where $Γ_k$ smoothly interpolates between the microscopic action $Γ_Λ$ and the full effective action $Γ_{k=0}=Γ$.
Each infinitesimal change of the average effective action is described by a flow equation $\partial_k\,Γ_k$ depending on the correlation function of the theory and a way how to suppress infrared modes with momenta smaller than $k$. In the end fluctuations with large wavelengths are included. Since the functional renormalisation group includes the fluctuations stepwise, there are no infrared divergences when approaching the inclusion of long wavelength modes. 
Analogous to defining the quantum theory by means of the classical action in the path integral formulation, the initial effective action $Γ_Λ$ together with the flow equation \eqref{eq:flow_eq_theo} determines the full quantum theory.

The infinitesimal change of the effective action $Γ_k$ with respect to the momentum scale $k$ is governed by the flow equation \cite{Wetterich:1989xg,Wetterich:1992yh,Pawlowski:2005xe,Schaefer:2006sr,Gies:2006wv,Delamotte:2007pf,Kopietz:2010zz,Metzner:2011cw,Braun:2011pp,Boettcher:2012cm,Dupuis:2020fhh}
\begin{align}\label{eq:flow_eq_theo}
	\partial_k\,Γ_k=\frac{1}{2}\,\text{STr}\left[(Γ^{(2)}_k+R_k)^{-1}\,\partial_k\,R_k\right]\,,
\end{align}
where $Γ^{(2)}_k$ is the second functional derivative of of $Γ_k$ with respect to the fields.
As the flow equation \eqref{eq:flow_eq_theo} is an integro-differential equation, its full solution is in most cases out of reach. One therefore relies on approximation schemes to the full effective action $Γ_k$ which should incorporate the examined physics already at lower order of the approximation and reduce the number of flow equations to a manageable set of couplings. Furthermore, it is convenient to rewrite the four-Fermi interaction $λ_ψ$ at a large cutoff $Λ$ in terms of a bosonic degree of freedom $ϕ$ via a Hubbard-Stratonovich transformation.

In this work we choose a three-dimensional Litim-type regulator \cite{2001IJMPA..16.2081L,Litim:2001up,PhysRevB.77.064504} for the cutoff function $R(Q)$ in three spatial dimensions. It is given for bosons and fermions, respectively, by
\begin{align}\label{eq:opt_reg}
\begin{split}
	&R_{ϕ,k}(q^2)= \left(k^2-q^2/2\right)\,θ\left(k^2-q^2/2\right)\,,\\[1ex]
	&R_{ψ,k}(q^2)= k^2\,\left(\sgn\left(z\right) -z\right)\,θ\left(1-|z|\right)\,,
	\end{split}
\end{align}
where $θ(x)$ represents the Heaviside-Theta function, $\text{sgn}(x)$ the sign function and we used $z=(q^2-μ)/k^2$. Note, that only spatial momenta $q^2=|\vec{q}|^2$ are regularised for this type of regulator. However, a particular neat property of \eqref{eq:opt_reg} is that the finite temperature Matsubara sums can be performed analytically.

\subsection{Function space and boundary conditions}\label{sec:BoundaryConditions}

The choice of the boundary conditions plays a crucial role in arriving at the correct two-dimensional physics.
The dimensional crossover is implemented by compactifying the
‘transverse’ $z$-direction by a potential well of length $L$, 
\begin{align}\label{eq:length_pot}
	V_{\text{box}}(z)=
	\begin{cases}
	0\qquad 0 \leq z \leq L\\[1ex]
	\infty \qquad \text{else}
	\end{cases}\,.
\end{align}
One may choose (anti-)periodic boundary conditions
\begin{align}
	ψ(x,y,z=0)=\pm ψ(x,y,z=L)\,,
\end{align}
or restrict oneself to a box
\begin{align}
  ψ(x,y,z=0)=ψ(x,y,z=L)=0\,.
\end{align}
The compactification leads to a discrete momentum
spectrum in $z$-direction. For periodic boundary conditions the
respective energies, ${E_z=\frac{\hbar\,q_z^2}{2\,M}}$, are discrete
with $q_z\rightarrow k_n$
\begin{align}
	k_n=\frac{2\,π\,n}{L},\qquad n\in \mathbb{Z}\,, 
\end{align}
which includes a zero mode $k_0=0$ with vanishing energy
${E_\textrm{min}=0}$. In turn, for anti-periodic boundary conditions one
finds ${k_n=(2\,n+1) \,π/{L}}$ with $n\in \mathbb{Z}$ and with a lowest
mode ${|k_0|=\pi /L}$ with a finite energy
${E_\textrm{min}= \hbar \pi^2/(2 M L^2)}$. Finally, confining the Fermi
gas inside a box leads to ${k_n=π\,n/L}$ with a vanishing energy
${E_\textrm{min}=0}$.

The non-vanishing zero point energy for anti-periodic boundary conditions results in a gap in the evaluation of the the (discrete) mode sum at zero temperature. Consequently, anti-periodic boundary conditions do not yield the two-dimensional limit for vanishing length $L \rightarrow 0$. 
For a relativistic system the dispersion relations allows one to identify the length of the potential well $L$ with the inverse temperature $1/T$ in the evaluation of the discrete mode sum at zero temperature. As a result, $T=0$ and $L=L_0$ gives the same result as $T=1/L_0$ and $L=0$, i.e. the zero length limit $L\rightarrow 0$ at zero temperature $T=0$ corresponds to the limit of infinite temperature $T \rightarrow \infty$ at zero length $L=0$.
For a non-relativistic system the situation is less simple, since the dispersion relation allows no clear mapping between the temperature and the length of the system.
Nevertheless, it is clear that anti-periodic boundary conditions do not admit a two-dimensional limit for $L\rightarrow 0$.

Here we choose periodic boundary conditions, which result in a two-dimensional limit for vanishing length $L\rightarrow 0$. 
Since all modes with $n≠0$ have for $L\rightarrow 0$ a large gap they can be integrated out. In general, the three-dimensional system with finite $L$ can be viewed as a two-dimensional system with infinitely many fermions as “modes”, one for each $n$. Integrating out the modes with $n≠0$ reduces the system to a single two-dimensional fermion, the one for $n=0$.

The map from the three-dimensional system to the two-dimensional system proceeds by integrating out the $n \neq 0$ modes. This maps the parameters of the three-dimensional theory to the ones of an effective two-dimensional theory. For $L \rightarrow 0$ this map may induce large changes for characteristic quantities as the chemical potential $μ$ or the scattering length $a$. This can lead to shifts in fractions including $ε_F$ and $T_F$, as well as in the crossover parameter. For experiment, the three-dimensional quantities are generally the ones available, and we will typically use them for our discussion. When comparing to results obtained from computations in two-dimensions, the matching between three- and two-dimensional parameters becomes important, however. In the present paper we do not deal with this issue, but the reader should keep it in mind when comparing with two-dimensional results.

Experimentally realistic confinement potentials, used in most ultracold atom experiment,
such as \cite{PhysRevLett.114.230401} and
\cite{2016PhRvL.116d5302F}, are implemented by using harmonic
trapping potentials.  Here, the function space consists of Hermite polynomials.
Heuristically, our choice is a limiting case.  In
particular, observables that do not show an impact of the different
boundary conditions studied here should be the same for the harmonic
trap.

\subsection{Dimensional reduction}\label{sec:DimRed}

In order to obtain a system within the dimensional crossover from three to two dimensions we initialise the renormalisation group (RG) flow at ultraviolet cutoff scale $k=Λ$ where the effective action $Γ_{Λ}$ coincides with the microscopic action of a three-dimensional ultracold Fermi gas. By delimiting the $z$-direction of the system via a potential well of length $L$ we introduce an additional scale to the three-dimensional system. By following the RG as a function of $k$ for a given length scale $L$ one observes that the contribution of modes with $k_n^2\gg k^2$ is suppressed by powers of $k^2/k_n^2$. These modes decouple and effective dimensional reduction is achieved automatically once $k \ll 2π/L$. This is very similar to the effective dimensional reduction in finite temperature quantum field theory realised by solutions of the flow equations \cite{Tetradis:1992xd}. Following the RG from $k=Λ$ to $k=0$ the flow always makes a transition from a three-dimensional regime to a two-dimensional one. For this purpose the UV scale is always chosen such that $Λ\gg (L^{-1},μ^{1/2},T^{1/2})$.
The flow equations become effectively two-dimensional for $k\ll 2π/L$, while the physical system is effectively two-dimensional if $L^{-1}$ is much larger than all other many-body scales \cite{2016PhRvA..93f3631L}.

To incorporate the effects of the compactification in transversal $z$-direction given by the potential well in \eqref{eq:length_pot}, the regulators in \eqref{eq:opt_reg} are modified according to
\begin{align}
	\vec{q}^2 = \hat{q}^2+q_z^2\rightarrow \hat{q}^2+k_n^2\,,
\end{align}
where $k_n$ is chosen according to the boundary conditions and $\hat{q}^2$ denotes the square of the $x$- and $y$-components of the momentum.
Note that while in three dimensions all couplings tend to saturate quickly at sufficiently small $k$-scales \cite{PhysRevA.89.053630}, the saturation behaviour is much slower for $d<3$. As a consequence, we choose a much smaller final $k$-scale in the infrared (cf. Appendix \ref{ap:IR-dep}), while in three dimensions it is possible to stop the RG-flow earlier.

\section{Running of couplings}\label{sec:runningCouplings}
\subsection{Truncation}

After a Hubbard-Stratonovich transformation the full microscopic action is given by
\begin{align}
\begin{split}
	S=\int_X &\left[\sum_{σ=1,2} ψ_σ^{*}\left(\partial_τ-\nabla^2-\overbar{μ}\right)ψ_σ\right.\\[1ex]
	&+m_ϕ^2\,ϕ^*ϕ-\left.h\,\left(ϕ^*\,ψ_1\,ψ_2-ϕ\,ψ_1^*\,ψ_2^*\right)\vphantom{\sum_{σ=1,2}}\right]\,,
\end{split}
\end{align}
with $\overbar{λ}_ψ=-h^2/m_ϕ^2$, which can be seen via a Gaussian integration over the bosonic field $ϕ$. The Feshbach coupling $h$ accounts for the interconversion of two fermionic atoms $ψ$ with different spin to a bosonic dimer $ϕ$.
Connecting the above action to the experimental setup we explicitly introduce the closed channel via the bosonic field $ϕ$. The physical detuning $ν=ν(B)$, which depends on the external magnetic field of the trap in the experiment, denotes the distance of the closed-channel bound state from the scattering threshold. In the kinetic term of the bosonic dimer $ϕ$ the factor of $\nabla^2/2$ reflects the composite mass of the dimer, while this composition also yields twice the chemical potential for the bosons
\begin{align}
\begin{split}
	S[ψ,ϕ]=\!\int_X\,&\left[\vphantom{\frac{\nabla^2}{2M}} ψ^{*}\left(\partial_τ-\nabla^2-μ\right)ψ\right.\\[1ex]
	\,&+\left.ϕ^*\left(\partial_τ-\frac{\nabla^2}{2}+ν-2μ\right)ϕ\right.\\[1ex]
	\,&-\left.h\,\left(ϕ^*\,ψ_1\,ψ_2-ϕ\,ψ_1^*\,ψ_2^*\right) \vphantom{\frac{\nabla^2}{2M}}\right]\,.
\end{split}
\end{align}
Our ansatz for the effective average action can be divided into a kinetic part and an interaction part
\begin{align}
	Γ_k = Γ_{\text{kin}}+Γ_{\text{int}}.
\end{align}
The kinetic part describes the fermion and boson dynamics and is given by
\begin{align}\label{eq:ansatz_kin_unren_fields}
  \begin{split}
  Γ_{\text{kin}}=\int_{X}\!\!\left[\sum_{σ=\{1,2\}}\!\!\!\overbar{ψ}^*_σ\,\overbar{P}_{ψ,σ}(Q)\,\overbar{ψ}_σ
  + \overbar{ϕ}^*\,\overbar{P}_{ϕ}(Q)\,\overbar{ϕ}\right]\,,
\end{split}
\end{align}
with unrenormalised (unrescaled) fields $\overbar{ψ}$, $\overbar{ϕ}$ and inverse propagators
\begin{align}\label{eq:propagator}
  \begin{split}
    \overbar{P}_{ψσ}(Q) &= Z_{ψσ}\,\iu\,q_0+A_{ψσ}\,q^2-\overbar{μ}\,,\\[1ex]
    \overbar{P}_{ϕ}(Q) &= Z_{ϕ}\,\iu\,q_0+A_{ϕ}\,q^2/2\,.
  \end{split}
\end{align}
In terms of the renormalised (rescaled) fields ${ψ=A_ψ^{1/2}\,\overbar{ψ}}$ and ${ϕ=A_ϕ^{1/2}\,\overbar{ϕ}}$ the kinetic part can be formulated as
\begin{align}\label{eq:ansatz_kin}
	\begin{split}
	Γ_{\text{kin}}[ψ,ϕ]=\int_{X} &\left[\sum_{σ=\{1,2\}}\!\!\!ψ^*_σ\left(S_ψ\,\partial_τ-\nabla^2-μ\right)\,ψ_σ\right.\\[1ex]
	&+ \left.ϕ^*\left(S_ϕ\,\partial_τ-\frac{1}{2}\nabla^2\right)\,ϕ\right]\,.
\end{split}
\end{align}
We normalised the coefficients of the gradient terms by means of the wave function renormalisations $A_ψ$ and $A_ϕ$ which enter the renormalisation group flow via the anomalous dimensions
\begin{align}
	η_ψ=-\,\partial_t \, \log{A}_ψ, \quad η_ϕ=-\,\partial_t \, \log{A}_ϕ\,.
\end{align}
Moreover, we defined $S_{ψ,ϕ}=Z_{ψ,ϕ}/A_{ψ,ϕ}$ and the renormalised chemical potential $μ=\overbar{μ}/A_{ψ,σ}$. For a more detailed description we refer to App. \ref{ap:floweq}.\\
Due to the renormalisation of the fields the expectation value $Δ=(h^2\,ρ)^{1/2}$ can be non-zero,
even in the two-dimensional limit \cite{2016PhRvA..93f3631L}, where the Mermin-Wagner theorem \cite{PhysRevLett.17.1133,PhysRev.158.383} forbids true long-range order.
However, algebraically decaying correlation functions with a non-vanishing superfluid density can be found \cite{Berezinsky:1970fr,1972JETP...34..610B,1973JPhC....6.1181K,1974JPhC....7.1046K}.

The interactions can, after the Hubbard-Stratonovich transformation, be written as
\begin{align}
	Γ_{\text{int}}[ψ,ϕ]=\int_{X} \, \left[U\left(ϕ^*\,ϕ\right)-h\,\left(ϕ^*\,ψ_1\,ψ_2 - ϕ\,ψ_1^*\,ψ_2^*\right)\vphantom{\frac{\nabla}{2m}}\right].
\end{align}
The effective average potential $U(ρ)$ depends only on the $U(1)$-invariant quantity $ρ=ϕ^*\,ϕ$ and describes bosonic scattering processes. The $U(1)$-symmetry is spontaneously broken for a non-zero minimum $ρ_0$ of the effective average potential and thus describes superfluidity.
In a Taylor-expansion we write
\begin{align}\label{eq:effpot}
	U(ρ)= m_ϕ^2\,\left(ρ-ρ_0\right)+\frac{λ_ϕ}{2}\left(ρ-ρ_0\right)^2+\sum_{n = 3}^N\frac{u_n}{n!}\left(ρ-ρ_0\right)^n.
\end{align}
where we need to include at least up to the second order in $ρ$ to reproduce the second order phase transition to superfluidity.
In the symmetric regime we therefore have $ρ_0=0$ and positive bosonic mass $m_ϕ^2>0$, whereas the symmetry-broken regime is realised for $ρ_0>0$ and vanishing bosonic mass $m_ϕ^2=0$. In the following we restrict this work to order $ϕ^4$.

\begin{figure}[t]
\centering
\includegraphics[width=8cm]{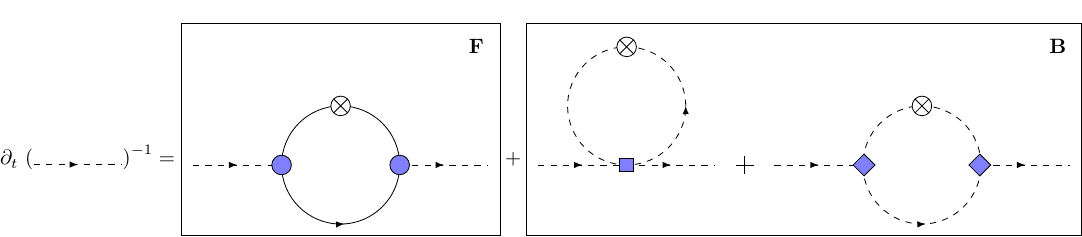}
\caption{(F)- and (B)-truncation schemes of the flow equations. The
  flow of the inverse boson propagator incorporates both fermionic and
  bosonic diagrams. Bosonic propagators correspond to dashed and
  fermionic propagators to solid lines, while the distinct vertices
  are shown in different shapes. The regulator insertion is denoted by
  a cross.}
\label{fig:feynman_diag_FB}
\end{figure}

The truncation can be classified by the diagrams in
Fig. \ref{fig:feynman_diag_FB} included on the right hand side of the
flow equation \eqref{eq:flow_eq_theo}. By including only fermionic diagrams (F) we arrive at
the mean-field result. Bosonic fluctuations enter the flow equation by including diagrams with two internal bosonic lines (B).

Furthermore, the flow of the density of the Fermi gas is calculated via a derivative
of the effective action with respect to the chemical potential
\begin{align}
	\partial_k\,n_k=-\partial_k\frac{\partial\, U(ρ)}{\partial\, μ}\,.
\end{align}
In practice, we approximate the dependence of the effective average action on the chemical potential by an
expansion in $ρ$ and $μ$ \cite{Diehl:2009ma}
\begin{align}
	U(ρ)&= \sum_{n = 1}^2\frac{u_n}{n!}\left(ρ-ρ_0\right)^n -n_k\,δμ+α_k\,(ρ-ρ_0)\,δμ\,.
\end{align}
Here the chemical potential is split into a reference part $μ_0$ and an offset $δμ$, such that $μ=μ_0+δμ$.

\subsection{Initial conditions and universality}

In three dimensions the running couplings approach
fixed points in the renormalisation group flow of the Fermi gas.
As a result, the macrophysics (on the length
scales of the inter-particle spacing) becomes independent of the
microphysics (on the molecular scales) to a large extent,
cf. e.g. \cite{Diehl:2009ma, PhysRevA.89.053630}.

When reaching the fixed points the system loses its memory of
the microphysics with its initial conditions. Consequently, the initial
conditions of the running couplings are irrelevant and we may essentially
start at the fixed point values in the ultraviolet.
Even if we had not done so, they would be immediately generated.

An exception constitutes the bosonic mass term $m_ϕ^2$ whose fixed
point is unstable towards the infrared. Hence, for the effective
potential we set as initial condition in the ultraviolet
\begin{align}
	U_Λ(ρ)=\left(ν_Λ-2\,μ\right)\,ρ\,.
\end{align}
Herein the chemical potential $μ$ can be artificially split into a
vacuum component $μ_{\text{v}}$ and a many-body contribution
$μ_{\text{mb}}$ such that the vacuum part $μ_{\text{v}}$ equals half the binding
energy of a bosonic dimer $ε_B/2$ in three spatial dimensions.
The detuning $ν_Λ$ is related to the physical detuning via an
appropriate vacuum renormalisation \cite{Diehl:2009ma}.

Since the RG flow for a system in reduced dimension is initialised
at an UV scale where the Fermi gas is described by the
three-dimensional classical action, these considerations can be applied
to the study of systems inside the dimensional crossover.
We therefore choose the fixed point values of the three-dimensional Fermi gas
as our initial conditions.

\begin{figure}[t]
\centering
\includegraphics[width=\linewidth]{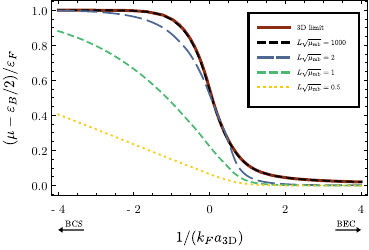}
\caption{Comparison of the equation of state at zero temperature for
  different confinement length scales and the three dimensional case
  with respect to the 3D crossover parameter
  $1/(k_F\,a_{\text{3D}})$ for $t_\text{final}=-17$. From top to bottom: 3D limit in solid-red (solid-grey),
  $L\,\sqrt{μ_{\text{mb}}}=1000$ in dashed-black,
  $L\,\sqrt{μ_{\text{mb}}}=2$ in long-dashed-blue (long-dashed line in dark grey),
  $L\,\sqrt{μ_{\text{mb}}}=1$ in dashed-green (dashed line in grey),
  $L\,\sqrt{μ_{\text{mb}}}=0.5$ in dotted-yellow (dotted line in light grey). The three-dimensional
  case is recovered for large $L\,\sqrt{μ_{\text{mb}}}$.}
\label{fig:dX_EqStateComp}
\end{figure}

\section{Dimensional crossover at zero temperature}\label{sec:zeroT}

The flow equations underlying the results at zero and at finite
temperature shown below are obtained analytically with periodic
boundary conditions for both bosonic and fermionic fields inside the
potential well. They are given in Appendix \ref{ap:floweq}.  Imposing
anti-periodic boundary conditions for fermionic fields $ψ(x)$ we find, 
as expected in Section \ref{sec:BoundaryConditions}, that for small
confinement length scales $L\sqrt{μ_{\text{mb}}}\sim 2$
the fermionic flow is strongly suppressed and no phase transition on
the BCS-side of the crossover can be found. The BEC-side, however, is
not affected by this choice.

\begin{figure}[t]
\centering
\includegraphics[width=\linewidth]{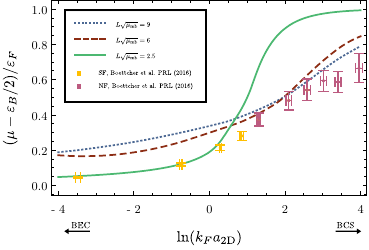}
\caption{Comparison of the equation of state for different confinement
  length scales to the experimental data from \cite{2016PhRvL.116d5303B}
  with respect to the 2D crossover parameter
  $\text{ln}(k_F\,a_{\text{2D}})$ for $t_\text{final}=-17$. Here we show:
  $L\,\sqrt{μ_{\text{mb}}}=9$ in dotted-blue (dotted line in grey), $L\,\sqrt{μ_{\text{mb}}}=6$
  in dashed-red (dashed line in dark grey), $L\,\sqrt{μ_{\text{mb}}}=2.5$ in solid-green (solid line in light grey). The experimental data
  is obtained for the lowest attainable temperatures of $T/T_F\approx 0.05$ on the BEC-side
  and $T/T_F\approx 0.1$ on the BCS-side. The orange and purple (light grey and dark grey) squares denote measurements in the
  superfluid and normal phase.}
\label{fig:dX_EqStateComp_2d}
\end{figure}

As described in Section \ref{sec:DimRed} all running couplings
saturate quickly in the infrared
for the three-dimensional BCS-BEC crossover, while we for $d<3$ we
chose for all observables a final scale of $t_\text{final}=-17$ in 
the infrared. It is chosen such that for a small (quasi-) two-dimensional system
of confinement length of $L\,\sqrt{μ_{\text{mb}}} = 0.7$
the maximum temperature for the BCS-BEC crossover converges, cf. 
Appendix \ref{ap:IR-dep} for a more detailed comparison on the final IR scale of 
the RG-flow.

In order to display the the confinement in transversal direction we
introduce the dimensionless length parameter
$L\,\sqrt{μ_{\text{mb}}}$ of the potential well, where ${μ_{\text{mb}}=μ-ε_B/2}$
denotes the chemical potential for the three-dimensional gas with half
the dimer binding energy $ε_B/2$ being subtracted.

At zero temperature a reduction of the dimensionless confinement
length parameter $L\,\sqrt{μ_{\text{mb}}}$ leads to an increased
density and thereby to an increased Fermi energy $ε_F=k_F^2$.
As a consequence the equation of state $(μ-ε_B/2)/ε_F$ in Figs. \ref{fig:dX_EqStateComp} and
\ref{fig:dX_EqStateComp_2d} is lowered for more confined systems.

Here the Fermi momentum 
is calculated using the three-dimensional definition $k_F=(3\,π^2\,n)^{1/3}$
as the initial condition for the flow of the density is explicitly given 
for a three-dimensional system. This means that the Fermi momentum $k_F$
of the \mbox{(quasi-)} two-dimensional system has to be calculated by using the
functional form given in the ultraviolet, where the reduced dimension enters via 
the flow of the density.

In Fig. \ref{fig:dX_EqStateComp} the equation of state is shown as a
function of the three-dimensional crossover parameter
\mbox{$c^{-1}=(k_{F}\,a_{3\text{D}})^{-1}$}, which can be interpreted as
the inverse concentration of the Fermi gas. For large confinement length
scales $L\,\sqrt{μ_{\text{mb}}}$ the three-dimensional result is
recovered, while the equation of state in dependence of the
transversal extension starts to saturate only at the order of
$L\,\sqrt{μ_{\text{mb}}}=10^{-4}$ for a two-dimensional limit.

\begin{figure}[t]
\centering
\includegraphics[width=\linewidth]{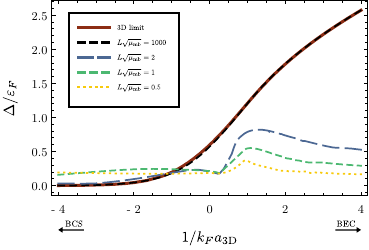}
\caption{Comparison of the gap parameter for different confinement
  length scales and the three dimensional case with respect to the 3D
  crossover parameter $1/(k_F\,a_{\text{3D}})$ for $t_\text{final}=-17$. From top to bottom (on BEC-side): 3D
  limit in solid-red (solid line in grey), $L\,\sqrt{μ_{\text{mb}}}=1000$ in dashed-black,
  $L\,\sqrt{μ_{\text{mb}}}=2$ in long-dashed-blue (long-dashed line in dark grey),
  $L\,\sqrt{μ_{\text{mb}}}=1$ in dashed-green (dashed line in grey),
  $L\,\sqrt{μ_{\text{mb}}}=0.5$ in dotted-yellow (dotted line in light grey). The three-dimensional
  case is recovered for large $L\,\sqrt{μ_{\text{mb}}}$.}
\label{fig:dX_GapComp}
\end{figure}

For better comparison to experiment the equation of state is also
displayed in Fig. \ref{fig:dX_EqStateComp_2d} with respect to the
two-dimensional crossover parameter
$\text{ln}\left(k_F\,a_{2\text{D}}\right)$. Here the
\mbox{(quasi-)} two-dimensional scattering length $a_{2\text{D}}$ is
calculated by \cite{2016PhRvA..93f3631L}
\begin{align}\label{eq:a2dpbc}
  a_{2\text{D}}^{(\text{pbc})}=L\,\exp\left\{-\frac{1}{2}\,
  \frac{L}{a_{3\text{D}}}\right\}
\end{align}
for our setup with periodic boundary conditions.  Comparing the result
in Fig. \ref{fig:dX_EqStateComp} with the experimental data found in
\cite{2016PhRvL.116d5303B} for a \mbox{(quasi-)} two-dimensional setup
we find a qualitatively good agreement.
Especially on the BEC-side, where the measurements 
were obtained in the superfluid phase, the equation of state for lower values of the confinement length $L\,\sqrt{μ_{\text{mb}}}$
our result yields the correct behaviour.
However, on the BCS-side the equation of state for confinements $L\,\sqrt{μ_{\text{mb}}}\lesssim 6$ does
not give the quantitative correct result.
This behaviour might be on the one hand attributed to an insufficient precision in the determination of the
density. For a more elaborate way to obtain the density see Appendix \ref{ap:mu-dep}.
On the other hand, as mentioned in Section \ref{sec:BoundaryConditions}, the two-dimensional limit for
periodic boundary conditions may feature parameters which do not coincide with the ones in three dimensions.

Comparing the gap parameter $Δ=(h^2\,ρ_0)^{1/2}$ with respect to the
Fermi energy $ε_F$ in Fig. \ref{fig:dX_GapComp} for different
confinement length scales one finds a flattening of the curve for lower
dimensionality, while the three-dimensional case is recovered for
large length scales $L\,\sqrt{μ_{\text{mb}}}$. Interestingly, the gap
saturates much faster for small length scales, already around
$L\,\sqrt{μ_{\text{mb}}}\simeq 0.5$ for a two-dimensional limit.
Moreover, depending on the (three-dimensional) scattering length $a_{\text{3D}}$, regions of an
increased gap $Δ/ε_F$ can be found at intermediate length scales within the dimensional crossover.
This dip-like structure is a characteristic of the modes given by the boundary conditions chosen and 
is also found at finite temperature.

\begin{figure}[t]
\centering
\includegraphics[width=\linewidth]{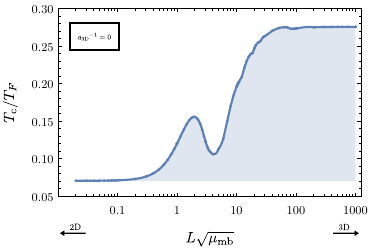}
\caption{Critical temperature $T_c/T_F$ as a function of the
  confinement length scale $L\,\sqrt{μ_{\text{mb}}}$ at an exemplary
  three dimensional fermion scattering length of
  $a_{3\text{D}}^{-1}=0$ for $t_\text{final}=-17$. Similar plots can be found for different
  scattering lengths with the difference being the amplitude and the
  position of the dips. These result from the mode structure caused by 
  the chosen boundary conditions and are related to the step-like structure of the density of states for a
  confined system. Similiar dips were also found in a mean-field analysis with 
  a harmonic confinement \cite{PhysRevB.90.214503}.}
\label{fig:Crossover_ainv_0}
\end{figure}

\section{Superfluid transition}\label{sec:dimXover}

\subsection{Dimensional crossover of the critical temperature}

At finite temperature we study the behaviour of the critical
temperature ${T_c/T_F}$ with respect to the spatial extension in transversal $z$-direction
$L\,\sqrt{μ_{\text{mb}}}$. The Fermi temperature
$T_F={k_F}^2$ is, as in the zero temperature case, 
calculated using the three-dimensional relation between the Fermi momentum
and the density $k_F=(3\,π^2\,n)^{1/3}$. The order parameter for the superfluid
transition is the (finite-temperature) gap $Δ=(h^2\,ρ)^{1/2}$ in the fermion spectrum.
The critical temperature is calculated as the largest temperature at which the gap $Δ$ is
non-vanishing. For a detailed description see App. \ref{ap:numerics}.

As shown exemplary for $a_{3\text{D}}^{-1}=0$ in
Fig. \ref{fig:Crossover_ainv_0} one can identify a dimensional
crossover from three to two dimensions for all values of the three-dimensional
scattering length. The limiting case of three dimensions is reached for large
confinement scales $L\,\sqrt{μ_{\text{mb}}}$. Moreover, a distinct
two-dimensional limit is obtained where the critical temperature in
units of the Fermi temperature saturates and is significantly reduced
with respect to the three-dimensional case. Note that as in the 
zero temperature case \ref{sec:zeroT} we choose $t_\text{final}=-17$ for the final RG-flow scale.

\begin{figure}[t]
\centering
\includegraphics[width=\linewidth]{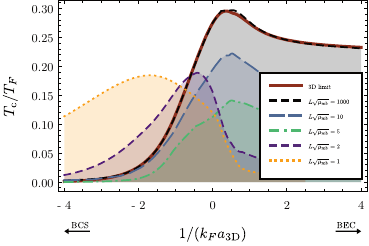}
\caption{Phase diagram in terms of $T_c/T_F$ for different confinement
  length scales and the three dimensional case with respect to the 3D
  crossover parameter $1/(k_F\,a_{\text{3D}})$ for $t_\text{final}=-17$. From top to bottom (on BEC-side): 3D
  limit in solid-red (solid line in grey), ${L\,\sqrt{μ_{\text{mb}}}=1000}$ in dashed-black,
  ${L\,\sqrt{μ_{\text{mb}}}=10}$ in long-dashed-blue (long-dashed line in grey),
  ${L\,\sqrt{μ_{\text{mb}}}=5}$ in dashed-dotted-green (dashed-dotted line in lighter grey), ${L\,\sqrt{μ_{\text{mb}}}=2}$ in dashed-purple (dashed line in dark grey) and
  ${L\,\sqrt{μ_{\text{mb}}}=1}$ in dotted-orange (dotted line in light grey).  }
\label{fig:PhaseDiagram3d}
\end{figure}

Furthermore, one can clearly discern dips in the dimensional crossover
of the critical temperature where we find an increased $T_c/T_F$ at
intermediate stages between the two- and three-dimensional
limit. Interestingly, their appearance and amplitude seem to be
related to the scattering length $a_{3\text{D}}$ chosen in the
ultraviolet. Moreover, we find a larger amplitude for more confined
systems. This behaviour is caused by the mode structure of a confined system
specified by the chosen boundary conditions. As a consequence, the density of states for a
confined system has a step-like structure and the dips can be found at 
the positions of the discontinuities. The dip structure for the critical temperature $T_c/T_F$
emerge at the same confinement length scales $L\,\sqrt{μ_{\text{mb}}}$ as for the zero temperature
gap parameter $Δ$. In a mean-field analysis with a confinement in transversal $z$-direction induced by a 
harmonic potential on the
weakly-interacting BCS-side of the BCS-BEC crossover a similar dip-like structure of the 
critical temperature was found \cite{PhysRevB.90.214503}.
The same behaviour is seen in
confined superconductors or thin superconducting films where the critical temperature, the gap parameter
and the intrapair correlation lengths are increased at so-called shape resonances 
\cite{Blatt_1963,Thompson_1963,Perali_1996,Innocenti_2010,Bianconi_2014,Pinto_2018,Eom_2006}.

\begin{figure}[t]
\centering
\includegraphics[width=\linewidth]{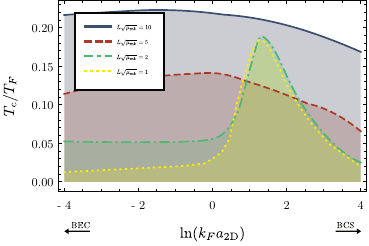}
\caption{Phase diagram in terms of $T_c/T_F$ for different confinement
  length scales with respect to the 2D crossover parameter
  $\text{ln}(k_F\,a_{2\text{D}})$ for $t_\text{final}=-17$. From top to bottom (on BEC-side):
  $L\,\sqrt{μ_{\text{mb}}}=10$ in solid-blue (solid line in dark grey), $L\,\sqrt{μ_{\text{mb}}}=5$
  in dashed-red (dashed line in grey), $L\,\sqrt{μ_{\text{mb}}}=2$ in dashed-dotted-green (dashed-dotted line in lighter grey) and
  $L\,\sqrt{μ_{\text{mb}}}=1$ in dotted-yellow (dotted line in light grey). The low critical temperature on
  the BEC-side is caused by our choice of boundary conditions, see Section
  \ref{sec:BoundaryConditions}.}
\label{fig:PhaseDiagram2d}
\end{figure}

\subsection{Finite temperature phase diagram}

In Figs. \ref{fig:PhaseDiagram3d} and \ref{fig:PhaseDiagram2d} the
critical temperature $T_c/T_F$ as a function of the three dimensional
inverse concentration $c^{-1}=(k_{F}\,a_{3\text{D}})^{-1}$
and the two-dimensional crossover parameter $\text{ln}(k_F\,a_{\text{2D}})$ 
is shown for different
confinement length scales over the whole BCS-BEC crossover. The phase
diagram in Fig. \ref{fig:PhaseDiagram3d} approaches the
three-dimensional limit for large confinement length scales, while the
critical temperature is reduced for lower dimensionality over the BCS-BEC crossover.
On the other hand, we find an increased
critical temperature on the BCS-side of the crossover around
$L\,\sqrt{μ_{\text{mb}}}=(0.5…5)$. On the BEC-side $T_c/T_F$ continues
to be reduced for more confined systems.

\begin{figure}[t]
\centering
\includegraphics[width=\linewidth]{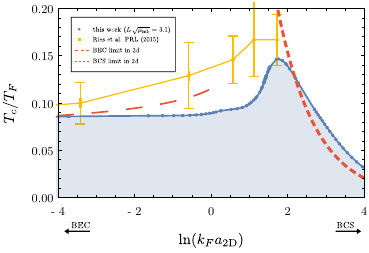}
\caption{Phase diagram in terms of $T_c/T_F$ for a confinement length
  of $L\,\sqrt{μ_{\text{mb}}}=3.1$ with respect to
  $\text{ln}(k_F\,a_{2\text{D}})$ for $t_\text{final}=-17$. Here we show the experimental data
  from \cite{PhysRevLett.114.230401} with the corresponding
  statistical errors in orange (light grey), as well as both the perturbative BKT-
  and BCS-transition temperatures as dashed red (dashed grey) lines in the
  appropriate regimes, i.e. ${\text{ln}(k_F\,a_{2\text{D}})\ll -1}$
  (BEC) and ${\text{ln}(k_F\,a_{2\text{D}})\gg 1}$ (BCS).}
\label{fig:PhaseDiagram2dComp2Point5}
\end{figure}

In Figs. \ref{fig:PhaseDiagram2d} and \ref{fig:PhaseDiagram2dComp2Point5} we
find the expected exponential decrease on the BCS-side of the
crossover, where $\text{ln}(k_F\,a_{2\text{D}})\gg 1$, for small confinement scales in a quasi-two-dimensional 
geometry. Here it was found \cite{10.1143/PTP.69.1794} that
\begin{align}
	\frac{T_c}{T_F}=\frac{2\,e^γ}{π\,k_F\,a_{2\text{D}}}
\end{align}
with the Euler number $γ\simeq 0.5772$. The critical temperature is
lowered by a factor of $e$ when including the Gorkov–Melik-Barkhudarov
contribution \cite{Gorkov:1961, PhysRevA.67.031601}.

Furthermore, the BKT-transition temperature on the BEC-side, where
$\text{ln}(k_F\,a_{2\text{D}})\ll 1$, is approximately reached for
these length scales. However, for smaller ${L\,\sqrt{μ_{\text{mb}}}}$, we obtain a
smaller value than the predicted BKT transition temperature
\cite{PhysRevA.67.031601,Levinsen:2014axa}
\begin{align}
  \frac{T_c}{T_F}=\frac{1}{2}\left[\log\left(\frac{
  \mathcal{B}}{4\,π}\,\log\left(\frac{4\,π}{k_F^2\,a_{2\text{D}}^2}\right)\right)\right]^{-1}\,,
\end{align}
with $\mathcal{B}\simeq 380$.

As described in Section \ref{sec:BoundaryConditions} 
this behaviour might be attributed to our choice of boundary conditions. Although
we are arriving at a two-dimensional system using periodic boundary conditions,
integrating out the higher modes in
the transversal $z$-direction may lead to a shift in the parameters of the Fermi gas.
This shift can also be differently pronounced depending on the scattering length.
The observation that $T_c/T_F$ decreases towards zero on the BEC-side for 
$L\rightarrow 0$ may be an indication for a strong $L$-dependence in the map from three-dimensional
to two-dimensional parameters in this region of the phase diagram and range of $L$.

\begin{figure}[t]
\centering
\includegraphics[width=\linewidth]{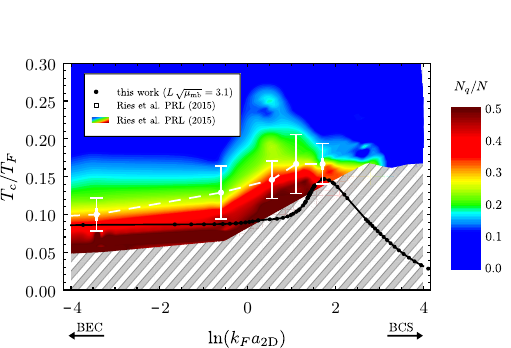}
\caption{Phase diagram in terms of $T_c/T_F$ for a confinement length
  of $L\,\sqrt{μ_{\text{mb}}}=3.1$ (in black) with respect to
  $\text{ln}(k_F\,a_{2\text{D}})$. Here we show the experimental data from \cite{PhysRevLett.114.230401}.
  The experimental critical temperature $T_c/T_F$ with the corresponding statistical errors is depicted in white, while
  the colour scale denotes the non-thermal fraction which signals the onset of a presuperfluid phase.}
\label{fig:PhaseDiagram2dCompNonThermal}
\end{figure}

In the region of strong correlations, where ${\text{ln}(k_F\,a_{2\text{D}})\simeq 1}$,
we find a substantial increase in
the critical temperature $T_c/T_F$ which cannot be found in a
mean-field analysis by extrapolation of the known BCS- and BEC-limits.

Comparing our results for $L\,\sqrt{μ_{\text{mb}}}=3.1$ to the
experimental data from \cite{PhysRevLett.114.230401} in
Fig. \ref{fig:PhaseDiagram2dComp2Point5}, where $L\,\sqrt{μ_{\text{mb}}}$ is
approximately of the order $0.5…5$, we find a qualitatively similar
phase diagram. Here the increased critical temperature in the strong
coupling regime can also be found, yet slightly less pronounced.

In Fig. \ref{fig:PhaseDiagram2dCompNonThermal} we show our result for 
a confinement length of $L\,\sqrt{μ_{\text{mb}}} = 3.1$ and the experimental
data on the non-thermal fraction found in \cite{PhysRevLett.114.230401}.
Here the preferred onset of a presuperfluid phase in the strongly correlated 
region is on par with our result of an increased superfluid temperature. The shift
with respect to the two-dimensional crossover parameter can be assigned to 
the change in the parameters from three to two dimensions of the Fermi gas.

\section{Conclusions and Outlook}\label{sec:conclusions}

In this paper we have studied the dimensional crossover in an
ultracold Fermi gas from three to two dimensions, thus extending the
work on non-relativistic bosons carried out in
\cite{2016PhRvA..93f3631L}, as well as the mean-field analysis in
\cite{PhysRevB.90.214503} for fermions. Particular emphasis was put on the
superfluid phase transition calculated over the whole BCS-BEC
crossover in dependence on different confinement length scales. A
comparison to recent experiments in \cite{2016PhRvL.116d5303B} and
\cite{PhysRevLett.114.230401} found a qualitative good agreement. 
Moreover, we find a non-trivial behaviour of the
finite temperature phase diagram when confining the Fermi gas in reduced dimensionality. Here
for small confinement length scales a substantial reduction of the
critical temperature $T_c/T_F$ on the one hand is found on the BEC-side of the
crossover, while on the other hand the critical temperature on the BCS-side is
moderately increased. Notably, in the strongly-coupled regime a
substantially higher critical temperature is found which is on
par with recent measurements \cite{PhysRevLett.114.230401}.

Within the dimensional crossover from
three to two dimensions a dip-like structure with regions of increased and reduced 
critical temperature $T_c/T_F$ were found. This dip-like structure is more or less pronounced
depending on the scattering length chosen in the ultraviolet
$a_{3\text{D}}$ and its exact shape is an artefact of the boundary conditions chosen for the 
confinement. For a harmonic confinement similar dips were seen in
\cite{PhysRevB.90.214503} for a mean-field study of the critical
temperature on the BCS-side for quasi-two dimensional Fermi gases.
Moreover, in confined superconducting systems this behaviour is known as superconducting
shape resonances and responsible for an increased critical temperature, gap and intrapair correlation length
at the discontinuities of the density of states.

These results suggest that a geometry lying between three and two
dimensions might be beneficial in finding systems with increased
critical temperature and thus in advancing in the quest for
high-$T_c$ superconductors.

Overall, we see that certain effects can be attributed to the 
dimensionality of the system. These include the dip-like structure
of increased and reduced critical temperature within the dimensional
crossover or the overall shape of the phase diagram at a certain confinement
length $L\,\sqrt{μ_{\text{mb}}}$. The effective dimension of the system has 
thus a constraining impact on the many-body physics.

The above procedure of confinement from three to two dimensions can in
general be extended to confinements from three to one and from two to
one dimensions, cf. e.g. \cite{Shanenko_2012} for a dimensional crossover from
two to one dimensions.  Moreover, for a more realistic confinement scenario a
harmonic trapping potential $V(z)=\frac{1}{2}m\,ω_z\,z^2$, as it is
approximately realised in most ultracold atom experiments, should be
implemented instead of the periodic conditions used in this
work in order to account for the correct trapping geometry. However,
already the periodic boundary conditions yield 
qualitatively similar features in the $L$-dependence of the critical
temperature as a harmonic trap.

A further quantitative improvement, within the dimensional crossover as
well as in three dimensions, concerns the calculation of the density by
which every quantity is normalised, by means of the Fermi momentum $k_F$.
As detailed in Appendix \ref{ap:mu-dep}, the initial conditions for observables
$g_i$ with scaling dimension $d_{g_i}\geq 2$ are dependent on the chemical potential $μ$.
As a consequence,  the flow of the density, calculated by an $μ$-derivative of the
effective potential, is not UV-finite. In Appendix \ref{ap:mu-dep} we outline
an iterative safe way of calculating the density whose results will be presented
in future work.
In addition, the truncation may be extended to include also the renormalisation
of the fermion propagator, as well as higher orders in the derivative expansion.

Another interesting aspect would be the study of spin- and mass-imbalanced Fermi gases
within the dimensional crossover, since here the influence of mismatching
Fermi surfaces and stronger fluctuations in lower dimensions might result in competing
effects concerning pairing \cite{He_2008,2012PhRvA..85e1607S,2015PhRvL.114k0403O,2016PhRvL.117i3601M,
2016PhRvA..94c3604W,2016PhRvA..94f3627D}.
This may shed further physical insight, for example in the search for high
temperature superconductors.

\begin{figure}[t]
\centering
\includegraphics[width=\linewidth]{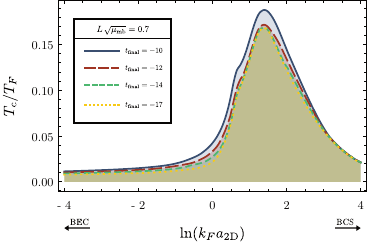}
\caption{Phase diagram in terms of $T_c/T_F$ for a confinement length
  of $L\,\sqrt{μ_{\text{mb}}}=0.7$. Here the dependence on the final IR scale of the RG-flow is shown with (from top to bottom) $t_\text{final}=-10$ in solid-blue (solid line dark grey), $t_\text{final}=-12$ in long-dashed-red (long-dashed line in grey), $t_\text{final}=-14$ in dashed-green (dashed line in lighter grey) and $t_\text{final}=-17$ in dotted-yellow (dotted line in light grey).
  It is most pronounced in the strongly-interacting region around $\ln{k_F\,a_\text{2D}} \sim 1$, while being much less significant in the BEC- and BCS-limits. A smaller IR-scale leads to a reduced critical temperature.}
\label{fig:PhaseDiagram_0Point7_tFinal_comp}
\end{figure}

Already at the present stage our beyond-mean-field analysis is an advancement
in the study of the interplay between many-body physics and dimensionality of ultracold Fermi gases.
It reveals that the dependence of fluctuation effects on the effective dimensionality leads
to new characteristic features that can be exploited in experiment and serve as a 
test for theoretical methods.

\section{Acknowledgements}

We thank Igor Boettcher, S\"oren Lammers, Stefan Fl\"orchinger, Selim
Jochim, Luca Bayha, Marvin Holten and Ralf Klemt for discussions.
We furthermore thank the group of Selim Jochim for providing their experimental
data to us.

The work is supported by EMMI and by the Deutsche Forschungsgemeinschaft 
(DFG, German Research Foundation) under the Collaborative Research Centre SFB 1225 (ISOQUANT) 
as well as under Germany’s Excellence Strategy EXC-2181/1-390900948 (the Heidelberg Excellence Cluster STRUCTURES).

\appendix

\section{Dependence on the IR RG-flow scale} \label{ap:IR-dep}

As mentioned in Section \ref{sec:DimRed} the running of the couplings does not saturate as quickly in $d<3$ as in three-dimensions in our RG-flow.
In Fig. \ref{fig:PhaseDiagram_0Point7_tFinal_comp} we show this IR-scale dependence exemplary for a confinement length of $L\sqrt{μ_{mb}}=0.7$ and find that it is most pronounced in the strongly interacting region around $\ln{\left(k_F\,a_\text{2D}\right)} \sim 1$ while being much less significant in the BEC- and BCS-limits.
Seeing that the critical temperature is reduced for a smaller IR-scale demands that we choose a sufficiently small final $k$-scale when solving our flow equations.

\begin{figure}[t]
\centering
\includegraphics[width=\linewidth]{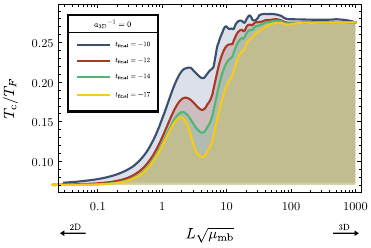}
\caption{Critical temperature $T_c/T_F$ at fixed three-dimensional scattering length $a_\text{3D}^{-1}=0$ over the dimensional crossover from two- to three-dimensions. Here we show the dependence on the final IR scale of the RG-flow with (from top to bottom) $t_\text{final}=-10$ in blue (dark grey), $t_\text{final}=-12$ in red (grey), $t_\text{final}=-14$ in green (lighter grey) and $t_\text{final}=-17$ in yellow (light grey).}
\label{fig:DimXover_tFinal_comp}
\end{figure}

The dependence on the final RG-flow scale in the infrared across the dimensional crossover can be seen in Fig. \ref{fig:DimXover_tFinal_comp} for a three-dimensional scattering length of $a_\text{3D}^{-1}=0$.
Again, the dips within the crossover from two- to three-dimensions are caused by the chosen boundary conditions and are related to the step-like structure of the density of states for a confined system \cite{PhysRevB.90.214503}.

\begin{figure}[t]
\centering
\includegraphics[width=\linewidth]{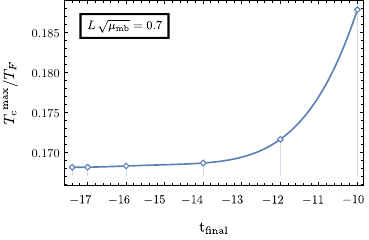}
\caption{Maximum critical temperature $T^{\text{max}}_c/T_F$ within the (quasi-) two-dimensional BCS-BEC crossover for a confinement length scale $L\,\sqrt{μ_{\text{mb}}}$ and different final $k$-scales in the infrared. The convergence for $t_{\text{final}}\leq-16$ can be inferred.}
\label{fig:tFinal_comp}
\end{figure}

As shown in Fig. \ref{fig:tFinal_comp} the maximum critical temperature $T^{\text{max}}_c/T_F$ within the (quasi-) two-dimensional BCS-BEC crossover for a confinment length scale $L\,\sqrt{μ_{\text{mb}}}=0.7$ converges for $t_{\text{final}}\leq-16$. For this reason we choose a final IR-scale of $t_{\text{final}}=-17$ for all our calculations such that sufficiently converged results should be obtained.

\section{\texorpdfstring{$\mu$-dependence}{Dependence on the chemical potential}}\label{ap:mu-dep}

In this appendix we discuss the potential $\mu$-dependences of initial conditions as well as an iterative safe way of how to extract related observables such as the density and higher $\mu$-derivatives of the
free energy. A similar procedure can be found in \cite{Fu:2015naa}.

It is well-known that thermal fluctuations decay exponentially with the infrared cut-off scale,
\begin{align}\label{eq:ex-decay}
  f(k/T,R) e^{ -c(R) k/T} \,,
\end{align}
where $f(k/T,R)$ rises not more than polynomially or even decays, depending on the (canonical) dimension of the observable under consideration, see \cite{Fister:2011uw}. The form of the prefactor as well as the coefficient $c(R)$ depend on the shape of the regulator. In particular, for non-analytic cut-offs (in frequency) such as the sharp cut-off and the optimal cut-off we have $c(R)=0$ and the thermal behaviour at large cut-off scales relates to the dimension of the observable. Note that \eq{eq:ex-decay} can be shown to hold to any order of a given approximation scheme and hence is a formal, exact property of thermal fluctuations. It is intimately linked to the fact that thermal sums can be represented as contour integrals and the infrared cut-off scale serves as a mass parameter which shifts poles
to momenta $p^2 \propto i k^2$. This also hints at the fact that it is not present for non-analytic regulators, where the Matsubara sum cannot be represented as a contour integral, and a naive dimensional analysis prevails.

In contradisctinction, the chemical potential $\mu$ as well as other external tuning parameters  only lead to a polynomial decay or rise in the dimensionless ratio
\begin{align}\label{eq:hatmu}
  \hat k=\0{k}{\mu}\,,\qquad \hat k=\0{k}{\sqrt{\mu}}\,,
\end{align}
for the relativistic case and non-relativistic case respectively. In most cases this behaviour is related to the (canonical) dimension of the observable at hand. For example, the free energy or effective action has a vanishing canonical dimension. However, it relates to (negative) pressure times space-time volume $\CV$ and hence has a scaling dimension $d_p=d$ with the cut-off scales in the relativistic
case and scaling dimension $d_p=d+1$ in the non-relativistic case.

The above arguments entail that the flow of the thermal pressure,
\begin{align}\label{eq:flowp}
  \partial_t p(T,\mu):= -\left(\0{\partial_t
  \Gamma_{k}[\phi_{{\rm EoS},k};T,\mu]}{\CV_T}-\0{\partial_t
  \Gamma_{k}[\phi_{{\rm EoS},k};0,\mu]}{\CV_0}\right)\,,
\end{align}
in general decays exponentially for large cut-off scales,
\begin{align}\label{eq:flowp-scaling}
  \partial_t p(T,\mu)\propto e^{ -c(R) k/T}\,,
\end{align}
while the free energy density, $f$, normalised in the vacuum,
\begin{align}\label{eq:flowf}
  \partial_t f(T,\mu):= \left(\0{\partial_t
  \Gamma_{k}[\phi_{{\rm EoS},k};T,\mu]}{\CV_T}-\0{\partial_t
  \Gamma_{k}[\phi_{{\rm EoS},k};0,0]}{\CV_0}\right)\,,
\end{align}
has polynomial growth with $k$,
\begin{align}\label{eq:flowf-scaling}
  \partial_t f(T,\mu)\to  c_{d_f-2}  k^{d_f} \hat k^{-2}+
  c_{d_f-4} k^{d_f}\hat k^{-4} + k^{d_f} O(\hat k^{-6})\,,
\end{align}
Here, vanishing exponents (in the relativistic case) include logarithms.

\subsection{Initial conditions}\label{sec:initial}
 Evidently, the initial conditions for observables or couplings $g_i$ with scaling dimension $d_{g_i}\geq 2$ are $\mu$-dependent. In turn, for sufficiently large cut-off scales $\hat k\gg 1$ the initial conditions for couplings with scaling dimension $d_{g_i} <2$ do not change when changing the chemical potential.

First we concentrate on the effective action, the flow of which is the master equation in our approach,
\begin{align}\label{eq:flow}
  \partial_t \Gamma_k[ψ,\phi]= \012 \Tr \,G_{k,ϕ} \,\partial_t R_{k,ϕ}
  - \Tr \,G_{k,\psi}\,\partial_t R_{k,\psi}\,,
\end{align}
where the field $\phi$ stands for bosonic fields while $\psi$ stands for fermionic ones. Every observable and coupling can be derived directly from \eq{eq:flow} and its solution. Indeed, if different definitions of observables such as the density exist, the one {\it directly} using the flow \eq{eq:flow} has the smallest systematic uncertainty.

For our investigation we write the effective action as
\begin{align}\label{eq:effac}
  \Gamma_k=\Gamma_k[ψ,\phi;\vec g],\quad
  \vec g=(m_\psi,m_\phi, Z_\psi, Z_\phi\,, h, \lambda_\psi, \lambda_\phi, ...)\,,
\end{align}
where $\vec g$ encodes all couplings (expansion coefficients) of the effective action, ordered in decaying mass dimension. We conclude that in $d=4$ dimensions the only couplings that potentially require $\mu$-dependent initial conditions are the mass parameters (including $\mu$ itself). However, the flow of the dimer mass reads asymptotically
\begin{align}
  \partial_t m_\phi^2\propto k  \0{h^2}{k} (1+\mu/k^2)^{3/2}
\end{align}
and hence its $\mu$-derivative tends towards zero, and the only coupling to be taken care of is the fermionic mass (and chemical potential).

\subsection{Density}\label{sec:density}
As already mentioned above, the equation for the density with the smallest systematic error is its flow. For the non-relativistic case it reads
\begin{align}\label{eq:flown}
  \partial_t n = \0{1}{\rm Vol}\0{d \partial_t \Gamma_k}{d\mu }
  \to c_{n,3} k^{3} + c_{n,1} \mu\, k
  + O(\hat k^{-1})\,,
\end{align}
and a similar equation holds for the relativistic case. The flow of the susceptibility reads
\begin{align}\label{eq:flowsus}
  \partial_t \partial_\mu n = \0{1}{\rm Vol}\0{d^2 \partial_t \Gamma_k}{d\mu^2 }
  \to c_{n,1} k + O(\hat k^{-1})\,,
\end{align}
while the flow of the second $\mu$-derivative of the density tends towards zero for large cut-off scales,
\begin{align}\label{eq:flowmusus}
  \partial_t \partial^2_\mu n = \0{\partial^3_\mu\partial_t \Gamma_k}{\rm Vol}
  \to O(\hat k^{-1})\,,
\end{align}
We conclude that we can represent the density, and the susceptibility at vanishing cut-off, $k=0$, by
\begin{align}\label{eq:nbyintegral}
   n(\mu) = \int_0^\mu d\mu'
  \partial_{\mu'}n(\mu')\,,
  \quad {\rm with} \quad
   n(0)=0\,,
\end{align}
and
\begin{align}\label{eq:susbyintegral}
  \partial_\mu n(\mu) = \int_0^\mu d\mu'\, \partial_{\mu'}^2n(\mu')\,,
  \quad {\rm with} \quad \partial_\mu n(0)=0\,,
\end{align}
It is left to determine $\partial_{\mu}^2n_k(\mu)$. To that end we rewrite the flow of the density as
\begin{align}\label{eq:flown-rewrite}
  \partial_t n_k=   \0{d \partial_t \Gamma_k}{d\mu } = \left.\partial_\mu \right|_{\vec g} \partial_t \Gamma_k
  +\0{d g_i }{d\mu }\partial_{g_i}\partial_t \Gamma_k\,.
\end{align}
Both terms follows analytically from the master equation, \eq{eq:flow}, and each partial $\mu$-derivatives and $d g_i/d\mu\, \partial_{g_i}$-derivative lowers the effective $k$-dimension by two. The coefficients $g_i^{ (1) }=d g_i/d\mu$ with
\begin{align}\label{eqmug}
  g^{(n)}_i=\0{d^n g_i}{d\mu^n}
\end{align}
follow from their flow
\begin{align}\label{eq:flowmul}
  \partial_t  g_i^{ (1) } = \0{d}{d\mu}\partial_t g_i=\partial_\mu\partial_t g_i  +
  g^{ (1) }_j\partial_{g_j}\partial_t g_i\,.
\end{align}
\Eq{eq:flowmul} is a coupled differential equation for $\vec g^{(1)}$,
\begin{align}\label{eq:flowmuform}
  \partial_t  \vec g^{ (1) }=\vec A_1 + B_1\cdot \vec  g^{ (1) }
\end{align}
with coefficients
\begin{align}\label{eq:coeffs1}
  A_{1,i}= \partial_\mu\partial_t g_i  \,,\quad B_{1,ij}= \partial_{g_j}\partial_t g_i\,.
\end{align}
The coefficients $A_{1,i}$ and $B_{1,ij}$ can be read-off from the flow \eq{eq:flow}, and hence \eq{eq:flowmuform} is a so-called derived flow: it does not feed back into the flow of the effective action. Naturally, this can be iteratively extended to the higher derivatives
w.r.t. $\mu$. For $g_i^{(2)}$ it reads
\begin{align}\label{eq:flowmul1}
  \begin{split}
    \partial_t  g_i^{ (2) } &=\frac{d}{d\mu}\left(A_{1,i}   +
    B_{1,ij}\, g^{ (1) }_j\right)\\[2ex]
    &= \partial_\mu A_{1,i} + g^{ (1) }_j\partial_{g_j}A_{1,i}\\[1ex]
    & + g^{ (1) }_j \left( \partial_\mu+ g_m^{ (1) }\partial_{g_m} \right)\,  B_{1,ij}+B_{1,ij}\, g^{ (2) }_j\,.
  \end{split}
\end{align}
Again this can be conveniently rewritten in terms of a system of linear differential equations
\begin{align}\label{eq:flowmu2}
  \partial_t  \vec g^{ (2) }=\vec A_2 + B_2\cdot \vec  g^{ (2) }\,,
\end{align}
with
\begin{align}\label{eq:coeffs2form}
\begin{split}
  A_{2,i}=\,\,& \left( \partial_\mu+ g_m^{ (1) }\partial_{g_m} \right)\, A_{1,i}\\[1ex]
  &+ g^{ (1) }_j \left( \partial_\mu+ g_m^{ (1) }\partial_{g_m} \right)\,  B_{1,ij} \,,\\[1ex]
  B_{2,ij}=\,\,& B_{1,ij}\,.
\end{split}
\end{align}
More explicitly we have
\begin{align}\nonumber
  A_{2,i}=\,\,& \partial_\mu^2\partial_t g_i + 2 g^{ (1) }_j\partial_{g_j}\partial_\mu\partial_t g_i  +
  g^{ (1) }_jg^{ (1) }_m\partial_{g_m} \partial_{g_j} \partial_t g_i \,,\\[1ex]
  B_{2,ij}=\,\,& \partial_{g_j}\partial_t g_i \,.
\label{eq:coeffs2}
\end{align}
This already allows us to put down the general structure. At a given order $g_i^{(n)}$ the matrix $B_n$ is simply $B_1$. The vector $A_n$ depends on $\vec g, \vec g^{(1)},...,\vec g^{(n-1)}$. Hence it can be determined iteratively with
\begin{align}\label{eq:Ait}
 \begin{split}
  A_{n,i}=& \left( \partial_\mu+
  \sum_{m=1}^{n-1} g_{j}^{(m)}
  \partial_{g^{(m-1)}_{j}}
  \right)\, A_{n-1,i}
  \\[2ex]
  & +
   g^{ (n-1) }_j \left( \partial_\mu+ g_m^{ (1) }\partial_{g_m} \right)\,  B_{ij}
 \end{split}
\end{align}
with $g_i^{(0)}=g_i$ and
\begin{align}\label{eq:Bder}
  \left( \partial_\mu+ g_m^{ (1) }\partial_{g_m} \right)\,  B_{ij} =
  \partial_\mu\partial_{g_j}\partial_t g_i  +
   g^{ (1) }_m\partial_{g_m} \partial_{g_j} \partial_t g_i \,.
\end{align}
For $n=3$ this explicitly yields
\begin{align}
  \begin{split}
  A_{3,i}=
  &\left[∂_{μ}^3+3\,g_j^{(1)}\,∂_{g_j}\,∂_μ^2+3\,g_j^{(1)}\,g_m^{(1)}\,∂_{g_m}\,∂_{g_j}\,∂_{μ}\right.\\[1ex]
  &+g_k^{(1)}\,g_j^{(1)}\,g_m^{(1)}\,∂_{g_m}\,∂_{g_j}\,∂_{g_k}\\
  &+\left.3\,g_m^{(2)}\,∂_{g_m}\,∂_μ +3\,g_j^{(2)}\,g_m^{(1)}\,∂_{g_m}\,∂_{g_j}\right]\,∂_t\,g_i\,.
  \end{split}
\end{align}
Note that there are various forms for the coefficients $A_n$ and $B_n$. The above forms have the advantage that all derivatives w.r.t. $\mu$ and $g^{(n)}_i$ can be performed analytically. Finally we write down the flow for higher $\mu$-derivatives of $\Gamma_k$
\begin{align}\label{eq:flowmoments}
  \partial_{\mu}^{(n-1)}\dot{n}(\mu)=
  \frac{d^{n}\partial_t \Gamma  }{d\mu^n}
  =  \left( \partial_\mu+\sum_{m=1}^{n} g_{j}^{(m)}\partial_{g^{(m-1)}_{j}}\right)\, C_{n-1}\,,
\end{align}
with
\begin{align}\label{eq:coefficientG}
  C_0=\partial_t\,\Gamma_k \,.
\end{align}
For $n=2$ this explicitly yields
\begin{align}
      \partial_{μ}\,∂_t\,{n}_k
      = \frac{d^2\,∂_t\,{Γ}_k}{dμ^2}
      =&\left[\left.\partial_μ^2\right|_{\vec g}
      +\,2\,g_i^{(1)}\,\partial_{g_i}\,\partial_μ\right.\\[1ex]
      &\left.+\,g_j^{(1)}\,g_i^{(1)}\,\partial_{g_i}\,\partial_{g_j}+g_i^{(2)}\,\partial_{g_i}
      \right]\,\partial_t\,Γ_k\,,\nonumber
\end{align}
while the second $μ$-derivative of the flow for the density is found to be
\begin{align}
  \begin{split}
    ∂_μ^2\,∂_t\,n_k=&\frac{d^3\,∂_t\,Γ_k}{dμ^3}\nonumber\\
    =&\left[∂_μ^3|_{\vec{g}}+3\,g_i^{(1)}\,∂_{g_i}\,∂_μ^2+3\,g_j^{(1)}\,g_i^{(1)}\,∂_{g_i}\,∂_{g_j}\,∂_μ\right.\\[1ex]
    &+g_m^{(1)}\,g_j^{(1)}\,g_i^{(1)}\,∂_{g_i}\,∂_{g_j}\,∂_{g_m}+3\,g_i^{(2)}\,∂_{g_i}\,∂_μ\\[1ex]
    &+\left.3\,g_i^{(2)}\,g_j^{(1)}\,∂_{g_j}\,∂_{g_i}+g_i^{(3)}\,∂_{g_i}\right]\,∂_t\,Γ_k\,.
  \end{split}
\end{align}
Hence, overall the density at vanishing cutoff $k=0$ is obtained by integrating twice over the chemical potential
\begin{align}
  \begin{split}
  n(μ)=&\int_0^μ\,dμ'\left[\int_0^{μ'}\,dμ''\,\partial_{μ''}^2\,n(μ'')+\partial_{μ'}\,n(0)\right]\\&+\,n(0)\,,
  \end{split}
\end{align}
where $n(0)$ and $\partial_μ\,n(0)$ are vanishing.

Moreover, we have
\begin{align}
  \partial_μ^2\,n_{k=0}(μ)=\int_Λ^0\,\frac{dk}{k}\,\partial_μ^2\,\dot{n}_k(μ)
\end{align}
for a UV vanishing flow $\partial_μ^2\,\dot{n}_{k\rightarrow \infty}\rightarrow0$.

\section{Flow equations}\label{ap:floweq}

In this appendix we derive the flow equations for an ultracold Fermi gas in the dimensional crossover. By defining a Master equation all flow equations of the individual couplings can be obtained by suitable projection descriptions.
Furthermore, we consider only the isotropic case where the flow of the couplings in transeversal direction equal the ones in the plane $g_i=g_{i,z}$, since this distinction is negligible \cite{2016PhRvA..93f3631L}. Our procedure is based on \cite{PhysRevA.89.053630}.

The ansatz for the effective average action can be divided in an kinetic part which consists of the fermion and boson dynamics and and interaction part
\begin{align}
	Γ_k = Γ_{\text{kin}}+Γ_{\text{int}}.
\end{align}
The kinetic part in terms of the renormalised fields $ψ=A_ψ^{1/2}\,\overbar{ψ}$ and $ϕ=A_ϕ^{1/2}\,\overbar{ϕ}$ is given by
\begin{align}\label{eq:ansatz_kin_2}
  \begin{split}
	Γ_{\text{kin}}[ψ,ϕ]=&\int_{X} \left[\sum_{σ=\{1,2\}}\,ψ^*_σ\left(S_ψ\,\partial_τ-\nabla^2+m_ψ\right)\,ψ_σ\right.\\[1ex]
	&+\left. ϕ^*\left(S_ϕ\,\partial_τ-V_ϕ\,\partial_τ^2-\nabla^2/2\right)\,ϕ\vphantom{\sum_{σ=\{1,2\}}}\right]\,.
  \end{split}
\end{align}
In this work we set $V_ϕ=0$. We normalised the coefficients of the gradient terms by means of the wave function renormalisations $A_ψ$ and $A_ϕ$ which enter the renormalisation group flow via the anomalous dimensions
\begin{align}
	η_ψ=-\,\partial_t \, \log{A}_ψ, \quad η_ϕ=-\,\partial_t \, \log{A}_ϕ.
\end{align}
Unrenormalised quantities are in the following denoted with an overbar, while renormalised ones are overbar-less.
The effective action $\overbar{Γ}[\overbar{ψ},\overbar{ϕ}]$ is expressed in terms of the unrenormalised fields. 
It can however be rescaled by appropriate powers of $A_{ψ,ϕ}$ such that $\overbar{Γ}[\overbar{ψ},\overbar{ϕ}]=Γ[ψ,ϕ]$.
The projections description is performed for the unrenomalised quantities, but the flow is evaluated for the renomalised ones.
The interactions can, after a Hubbard-Stratonovich transformation, be written as
\begin{align}
	Γ_{\text{int}}[ψ,ϕ]=\int_{X} \, \left(U\left(ϕ^*\,ϕ\right)-h\,\left(ϕ^*\,ψ_1\,ψ_2 - ϕ\,ψ_1^*\,ψ_2^*\right)\right)
\end{align}
neglecting the RG-flow of the four-fermion vertex $λ_{ψ,k}$.

The effective average potential depends only on the $U(1)$-invariant quantity $ρ=ϕ^*\,ϕ$ and describes bosonic scattering processes. The $U(1)$-symmetry is spontaneously broken for a non-zero minimum $ρ_0$ of the effective average potential and thus describes superfluidity.

In a Taylor-expansion we write
\begin{align}\label{eq:effpot_2}
	\begin{split}
		U(ρ)=\, & m_ϕ^2\,\left(ρ-ρ_0\right)+\frac{λ_ϕ}{2}\left(ρ-ρ_0\right)^2\\[1ex]
		&-n_k\,δμ+α_k\,(ρ-ρ_0)\,δμ\,,
	\end{split}
\end{align}
where we need to include at least up to the second order in $ρ$ to reproduce the second order phase transition to superfluidity.
In the symmetric regime we therefore have $ρ_0=0$ and $m_ϕ^2>0$, whereas the symmetry-broken regime is realised for $ρ_0>0$ and $m_ϕ^2=0$.

\subsection{Truncation}

By including only the fermionic diagrams (F) of Fig. \ref{fig:feynman_diag_FB} we arrive at the mean-field result and the bosonic fluctuations are taken care of by the diagrams including two bosonic lines (B).

The inverse propagators $\overbar{G}_ϕ^{-1}(Q)$ and $\overbar{G}_ψ^{-1}(Q)$ are calculated by
\begin{align}
\begin{split}
	&\overbar{Γ}^{(2)}_{\overbar{ϕ}_i,\overbar{ϕ}_j}(X,Y,\overbar{ρ})=\frac{δ^2\,\overbar{Γ}}{δ\overbar{ϕ}_i(X)\,δ\overbar{ϕ}_j(Y)}[\overbar{ϕ}]\\[1ex]
	&\overbar{Γ}_{\overbar{ψ}_α^{(*)},\overbar{ψ}_β^{(*)}}^{(2)}(X,Y,\overbar{ρ})=\frac{\overrightarrow{δ}}{δ\overbar{ψ}_α^{(*)}(X)}\,\overbar{Γ}\,\frac{\overleftarrow{δ}}{δ\overbar{ψ}_β^{(*)}(Y)}\,[\overbar{ϕ}]
\end{split}
\end{align}
where the boson background field $ϕ$ is assumed to be real valued and the direction of the arrow for the inverse fermion propagator denotes derivatives acting from left and right on the effective potential. In momentum space we arrive at
\begin{align}
\begin{split}
	\overbar{Γ}_{BB}^{(2)}(Q,Q')=δ(Q+Q')\,\overbar{G}_ϕ^{-1}(Q)\,,\\[1ex]
	\overbar{Γ}_{FF}^{(2)}(Q,Q')=δ(Q+Q')\,\overbar{G}_ψ^{-1}(Q)
\end{split}
\end{align}
After performing the functional derivatives we obtain in the $\{ϕ_1,ϕ_2\}$-basis for a constant bosonic background field $ϕ=\sqrt{ρ}$
\begin{align}
\begin{split}
	&\overbar{G}_ϕ^{-1}(Q)=A_ϕ\,
	\begin{pmatrix}
		P_ϕ^{S,Q}+U'+2\,ρ\,U'' & \iu\,P_ϕ^{A,Q}\\[1ex]
		-\iu\,P_ϕ^{A,Q} & P_ϕ^{S,Q}+U'
	\end{pmatrix}\\[1ex]
	&\overbar{G}_ψ^{-1}(Q)=A_ψ\,
	\begin{pmatrix}
		-h\,\sqrt{ρ}\,ε & -P_ψ^{-Q}\,\mathbbm{1} \\[1ex]
		P_ψ^{Q}\,\mathbbm{1} & h\,\sqrt{ρ}\,ε
	\end{pmatrix}
\end{split}
\end{align}
with $=\overbar{G}_Ψ^{-1}=A_ϕ\,G_Ψ^{-1}(Q)$, $\mathbbm{1}$ being the 2-dimensional unity matrix, $ε=((0,1),(-1,0))$ the fully antisymmetric tensor and a prime denotes a derivative with respect to $ρ$.

The regulators in the $\{ϕ_1,ϕ_2\}$-basis are given by
\begin{align}
\begin{split}
	&\overbar{R}_ϕ^Q=A_ϕ\,R_ϕ^Q=A_ϕ
	\begin{pmatrix}
		R_ϕ^S(Q) & \iu\, R_ϕ^A(Q)\\[1ex]
		-\iu\,R_ϕ^A(Q) & R_ϕ^S(Q)
	\end{pmatrix}\\[1ex]
	&\overbar{R}_ψ^Q=A_ψ\,R_ψ^Q=A_ψ
	\begin{pmatrix}
		0 & -R_ψ^{-Q} \mathbbm{1}\\[1ex]
		R_ψ^Q\,\mathbbm{1} & 0
	\end{pmatrix}
\end{split}
\end{align}
Moreover we defined the symmetrised and anti-symmetrised components of the propagators and regulator functions as
\begin{align}\label{eq:sym_unsym}
	f^{S,A}(Q)=\frac{f(Q)\pm f(-Q)}{2}.
\end{align}
By introducing short-hand notations for the sum of propagator and regulator, as well as the determinants
\begin{align}
\begin{split}
	L_ψ^Q&=P_ψ^Q+R_ψ^Q\\[1ex]
	\text{det}_F^Q&=L_ψ^Q\,L_ψ^{-Q}+h^2\,ρ\\[1ex]
	L_ϕ^Q&=P_ϕ^Q+R_ϕ^Q+U'+ρ\,U''\\[1ex]
	\tilde{L}_ϕ^Q&=P_ϕ^Q+R_ϕ^Q\\[1ex]
	\text{det}_B^Q&=L_ϕ^Q\,L_ϕ^{-Q}-(ρU'')^2.
\end{split}
\end{align}
we may write the regularised propagators as
\begin{align}
\begin{split}
	&G_ϕ^Q=A_ϕ\,\overbar{G}_ϕ^Q = \frac{1}{\detB}
	\begin{pmatrix}
		\tilde{L}_ϕ^{S,Q}+U' & -\iu\,\tilde{L}_ϕ^{A,Q}\\[1ex]
		\iu\,\tilde{L}_ϕ^{A,Q} & \tilde{L}_ϕ^{S,Q}+U'+2\,ρ\,U''
	\end{pmatrix}\\[1ex]
	&G_ψ^Q=A_ψ\,\overbar{G}_ψ^Q = \frac{1}{\detF}
	\begin{pmatrix}
		(h^2\,ρ)^{1/2}\,ε & L_ψ^{-Q}\,\mathbbm{1}\\[1ex]
		-L_ψ^{Q}\,\mathbbm{1} & -(h^2\,ρ)^{1/2}\,ε
	\end{pmatrix}
\end{split}
\end{align}
We can also represent the boson propagator in the conjugate field basis $\{ϕ,ϕ^*\}$ where the corresponding matrix will be labeled by a hat.\\
For $ϕ=(ϕ_1+\iu\,ϕ_2)/\sqrt{2}$ we have
\begin{align}
	\begin{pmatrix}
		ϕ\\[1ex]
		ϕ^*
	\end{pmatrix}
	 = \frac{1}{\sqrt{2}}
	\begin{pmatrix}
		1 &  \,\,\,\,\iu\\[1ex]
		1 &  -\iu
	\end{pmatrix}
	\begin{pmatrix}
		ϕ_1\\
		ϕ_2
	\end{pmatrix}
\end{align}
and thus arrive at
\begin{align}
	\widehat{G}_ϕ^{-1}= U\,G_ϕ^{-1}\,U^t
\end{align}
with the definitions
\begin{align}
	U = \frac{1}{\sqrt{2}}
	\begin{pmatrix}
		1 &  -\iu\\[1ex]
		1 & \,\,\,\, \iu
	\end{pmatrix}, \quad
	U^t = \frac{1}{\sqrt{2}}
	\begin{pmatrix}
		\,\,\, 1 &  1\\[1ex]
		-\iu &  \iu
	\end{pmatrix}\,.
\end{align}
Thus we obtain for the inverse boson propagator in the $\{ϕ,ϕ^*\}$-basis
\begin{align}
	\widehat{G}_ϕ^{-1}=
	\begin{pmatrix}
		ρ\,U'' & L_ϕ^{-Q}\\[1ex]
		L_ϕ^Q & ρ\,U''
	\end{pmatrix}, \quad
	\widehat{R}_ϕ(Q) =
	\begin{pmatrix}
		0 & R_ϕ^{-Q}\\[1ex]
		R_ϕ^Q & 0
	\end{pmatrix}
\end{align}
and
\begin{align}
	\widehat{G}_ϕ^{Q} = \frac{1}{\detB}
	\begin{pmatrix}
		-ρ\,U'' & L_ϕ^{-Q}\\[1ex]
		L_ϕ^Q & -ρ\,U''
	\end{pmatrix}
\end{align}
To generate higher n-point functions further functional derivatives have to be applied, once again paying attention to the correct ordering for fermionic derivatives.\\
Since we assume momentum and frequency independent vertices to close our set of equation, the complexity of the system of differential equations is drastically reduced
\begin{align}
	\overbar{Γ}_k^{(n>2)}(Q_1,…,Q_n)=\overbar{γ}_k^{(n)}\,δ(Q_1,…,Q_n)\,.
\end{align}
\vspace{.2cm}

\subsection{Master equations}
In order to solve the Wetterich equation in practice we need to convert it into a set of coupled differential equations of the correlation functions. We therefore start from a few Master equations, namely for the inverse fermion and boson propagators, the effective average potential and the Feshbach coupling.\\
In the next step these equation are projected appropriately to arrive at flow equations for the running couplings $\{g_k\}$.

For general regulators the flow equation of the effective average potential is then given by
\begin{align}\label{eq:master_effpot}
\begin{split}
	\dot{\overbar{U}}_k(\overbar{ρ})&=\frac{1}{2}\text{Tr}\int_Q\overbar{G}_ϕ^Q\dot{\overbar{R}}_ϕ^Q
	-\frac{1}{2}\text{Tr}\int_Q\overbar{G}_ψ^Q\dot{\overbar{R}}_ψ^Q\\[1ex]
	&=\frac{1}{2}\int_Q\frac{1}{A_ϕ}\frac{L_ϕ^Q\,\dot{\overbar{R}}_ϕ^{-Q}+L_ϕ^{-Q}\,\dot{\overbar{R}}_ϕ^Q}{\text{det}_B^Q}\\[1ex]
  & -\frac{1}{2}\int_Q\frac{1}{A_ψ}\frac{L_ψ^Q\,\dot{\overbar{R}}_ψ^{-Q}+L_ψ^{-Q}\,\dot{\overbar{R}}_ψ^Q}{\text{det}_F^Q}
\end{split}
\end{align}
Our flow equations can be divided into a bosonic and a fermionic contribution resulting from bosonic (B) and fermionic (F) diagrams, respectively.
\begin{align}
	\dot{\overbar{U}}(\overbar{ρ}) = \dot{U}^{(B)} + \dot{U}^{(F)}
\end{align}
Including the additional term of the anomalous dimension we find the flow for the renormalised quantities, e.g.
\begin{align}
	\dot{U}(ρ) = \dot{U}^{(B)} + \dot{U}^{(F)} + η_ϕ\,ρ\,U'(ρ).
\end{align}
\begin{widetext}
For the flow of the inverse boson propagator in the $\{ϕ_1,ϕ_2\}$-basis we find
\begin{align}
\begin{split}
	\dot{\overbar{G}}^{-1}_{\overbar{ϕ}_i\overbar{ϕ}_j}(P) =\,
	&\frac{1}{2}\,\tr\int_Q \overbar{G}_ϕ(Q)\,\overbar{γ}^{(3)}_{\overbar{ϕ}_iBB}\,\overbar{G}_ϕ(Q+P)\,\overbar{γ}^{(3)}_{\overbar{ϕ}_jBB}\,\overbar{G}_ϕ(Q)\dot{\overbar{R}}_ϕ(Q)\\[1ex]
	+&\frac{1}{2}\,\tr\int_Q \overbar{G}_ϕ(Q)\,\overbar{γ}^{(3)}_{\overbar{ϕ}_jBB}\,\overbar{G}_ϕ(Q-P)\,\overbar{γ}^{(3)}_{\overbar{ϕ}_iBB}\,\overbar{G}_ϕ(Q)\dot{\overbar{R}}_ϕ(Q)\\[1ex]
	-&\frac{1}{2}\,\tr\int_Q \overbar{G}_ϕ(Q)\,\overbar{γ}^{(4)}_{\overbar{ϕ}_i\overbar{ϕ}_jBB}\,\overbar{G}_ϕ(Q)\\[1ex]
	-&\frac{1}{2}\,\tr\int_Q \overbar{G}_ψ(Q)\,\overbar{γ}^{(3)}_{\overbar{ϕ}_iF|F}\,\overbar{G}_ψ(Q+P)\,\overbar{γ}^{(3)}_{\overbar{ϕ}_jF|F}\,\overbar{G}_ψ(Q)\dot{\overbar{R}}_ψ(Q)\\[1ex]
	-&\frac{1}{2}\,\tr\int_Q \overbar{G}_ψ(Q)\,\overbar{γ}^{(3)}_{\overbar{ϕ}_jF|F}\,\overbar{G}_ψ(Q-P)\,\overbar{γ}^{(3)}_{\overbar{ϕ}_iF|F}\,\overbar{G}_ψ(Q)\dot{\overbar{R}}_ψ(Q).
\end{split}
\end{align}
Likewise the flow of the inverse fermion propagator is obtained, taking the Grassmannian nature of fermions in account,
\begin{align}
\begin{split}
	\dot{\overbar{G}}^{-1}_{\overbar{ψ}_α\overbar{ψ}_β}(P) =\,
	&\frac{1}{2}\,\tr\int_Q \overbar{G}_ϕ(Q)\,\overbar{γ}^{(3)}_{\overbar{ψ}_αB|F}\,\overbar{G}_ψ(Q+P)\,\overbar{γ}^{(3)}_{F|B\overbar{ψ}_β}\,\overbar{G}_ϕ(Q)\dot{\overbar{R}}_ϕ(Q)\\[1ex]
	-&\frac{1}{2}\,\tr\int_Q \overbar{G}_ϕ(Q)\,\overbar{γ}^{(3)}_{BF|\overbar{ψ}_β}\,\overbar{G}_ψ(Q-P)\,\overbar{γ}^{(3)}_{\overbar{ψ}_α|FB}\,\overbar{G}_ϕ(Q)\dot{\overbar{R}}_ϕ(Q)\\[1ex]
	-&\frac{1}{2}\,\tr\int_Q \overbar{G}_ψ(Q)\,\overbar{γ}^{(3)}_{\overbar{ψ}_α|FB}\,\overbar{G}_ϕ(Q+P)\,\overbar{γ}^{(3)}_{BF|\overbar{ψ}_ϕ}\,\overbar{G}_ψ(Q)\dot{\overbar{R}}_ψ(Q)\\[1ex]
	+&\frac{1}{2}\,\tr\int_Q \overbar{G}_ψ(Q)\,\overbar{γ}^{(3)}_{F|B\overbar{ψ}_β}\,\overbar{G}_ϕ(Q-P)\,\overbar{γ}^{(3)}_{\overbar{ψ}_αB|F}\,\overbar{G}_ψ(Q)\dot{\overbar{R}}_ψ(Q).
\end{split}
\end{align}
\end{widetext}

\subsection{Projection description for the running couplings}
In this section we derive suitable projection descriptions for the flow equations of the running couplings $\{g_k\}$ and expansion coefficients of the effective average potential $U(ρ)$.
We use a derivative expansion of the inverse fermion and boson propagators
\begin{align}\label{eq:derivative_expansion_ren}
	\begin{split}
		\overbar{P}_{ψσ}(Q) &= Z_{ψσ}\,\iu\,q_0+A_{ψσ}\,q^2-\overbar{μ}\,,\\[1ex]
		&= A_{ψσ}\left(S_{ψσ}\,\iu\,q_0+q^2-μ
		\right)\\[1ex]
		\overbar{P}_{ϕ}(Q) & = Z_{ϕ}\,\iu\,q_0+A_{ϕ}\,q^2/2\\[1ex]
    &= A_{ϕ}\left(S_{ϕ}\,\iu\,q_0+q^2/2\right)\,.
	\end{split}
\end{align}
Here we defined $S_{ψ,ϕ}=Z_{ψ,ϕ}/A_{ψ,ϕ}$ and the renormalised chemical potential $μ=\overbar{μ}/A_{ψ,σ}$.
Expanding the effective potential in a Taylor series we can easily project the flow equation (\ref{eq:master_effpot}) onto the coefficients
\begin{align}
	U_k(ρ) = m_ϕ^2\,\left(ρ-ρ_0\right)+\frac{λ_ϕ}{2}\,\left(ρ-ρ_0\right)^2+\sum_{n>2}^N\frac{u_n}{n!}\,\left(ρ-ρ_0\right)^n.
\end{align}
There are several candidates for projection descriptions for the running couplings which may at a first glance seem equal. However, as the Wetterich equation is an exact equation incorporating all orders of the effective average action, every projection neglects certain higher order couplings and thus resultes in different flows. We expect though that our truncation includes the most important effects and a precise projection would only yield negligible modifications. The distinction between different projection descriptions may also be used for an error estimate.

In the symmetric regime of the flow we have $\dot{\overbar{m}}_ϕ^2=\dot{\overbar{U}}'(ρ=0)$ which makes place for the flow of $\dot{\overbar{ρ}}_0 = -\dot{\overbar{U}}'(ρ_0)/\overbar{λ}_ϕ$ in the symmetry broken regime. For the flow of higher expansion coefficients one finds
\begin{align}
	\dot{\overbar{u}}_n = \partial_t\,\left(\overbar{U}^{(n)}(\overbar{ρ}_0)\right) = \dot{\overbar{U}}^{(n)}(\overbar{ρ}_0) + \overbar{u}_{n+1}\,\dot{\overbar{ρ}}_0.
\end{align}
We obtain the flow of the renormalised couplings
\begin{align}
\begin{split}
	m_ϕ^2=\frac{\overbar{m}_ϕ^2}{A_ϕ},\quad
	ρ_0 = A_ϕ\,\overbar{ρ}_0,\quad
	u_n = \frac{\overbar{u}_n}{A_ϕ^n}
\end{split}
\end{align}
by
\begin{align}
\begin{split}
	&\dot{m}_ϕ^2 = η_ϕ\,m_ϕ^2+\frac{\dot{\overbar{m}}_ϕ^2}{A_ϕ}\,,\\[1ex]
	&\dot{ρ}_0 = - η_ϕ\,ρ_0 + A_ϕ\,\dot{\overbar{ρ}}_0\,,\\[1ex]
	&\dot{u}_n = η_ϕ\,u_n\,n + \frac{\dot{\overbar{u}}_n}{A_ϕ^n}\,.
\end{split}
\end{align}
Since we restrict ourselves to purely fermionic and bosonic diagrams, we have no running of the couplings entering the fermionic propagator.

For the couplings associated with the boson propagator we obtain
\begin{align}\label{eq:proj_B}
\begin{split}
	\dot{Z}_ϕ&=-\partial_{p_0}\left.\dot{\overbar{G}}_{ϕ_1ϕ_2}^{-1}(P,ρ_0)\right\vert_{P=0,ρ_0}\,,\\[1ex]
	\dot{A}_ϕ&=2\,\partial_{p^2}\left.\dot{\overbar{G}}_{ϕ_2ϕ_2}^{-1}(P,ρ_0)\right\vert_{P=0,ρ_0}\,,
\end{split}
\end{align}
and for the renormalised quantities with the anomalous dimension $η_ϕ=-\dot{A}_ϕ/A_ϕ$
\begin{align}
\begin{split}
	\dot{S}_ϕ=η_ϕ\,S_ϕ+\frac{\dot{Z}_ϕ}{A_ϕ}\,.
\end{split}
\end{align}
In the flow equations for the running couplings we neglected a term proportional to $\dot{\overbar{ρ}}_0$ which would be generated if one took the RG-time derivative after performing the projections.

\subsection{Flow equations using the optimised regulator}\label{sec:Opt_reg_flow}
In this section we use the optimised regulator \eqref{eq:opt_reg} for deriving the flow equations of the running couplings. These equations will be our main starting point in studying the BCS-BEC crossover in dimensions $2 \leq	 d \leq 3$.
The advantage of the optimised regulator stems from the possibility of analytically performing the Matsubara summations due to a purely spatial cutoff $q^2=|\vec{q}|^2$.\\
The procedure may, however, further be simplified by interchanging the order of the derivative projection and the Matsubara summation. We therefore start again from the general form of the flow of the inverse propagators with the trace not being evaluated so that we can expand the inverse propagators $G(Q\pm P)$ in powers of $p_0$ and $p$ and perform the projections afterwards.

For the fermionic contributions we arive at the general flow equations with the loop integration still unevaluated
\begin{align}
	\begin{split}
		\dot{S}_ϕ^{(F)} &= -2\,h^2\,S_ψ\,\int_Q\,\frac{\dot{\overbar{R}}_ψ(q^2)}{A_ψ}\left(\frac{1}{\detFsim^2}-\frac{2\,h^2\,ρ}{\detFsim^3}\right)\,,\\[1ex]
		η_ϕ^{(F)} &= \frac{8\,h^2}{d}\int_Q\,\frac{\dot{\overbar{R}}_ψ(q^2)}{A_ψ}\,\frac{q^2\,\rFtwo}{\detFsim^3}\,.
	\end{split}
\end{align}
The expansion for the bosonic contributions results in the flow equations
\begin{widetext}
\begin{align}\label{eq:flow_Sphi_eta}
\begin{split}
	\dot{S}_ϕ^{(B)} =&\, -\,4\,S_ϕ\,ρ\,U''\,\int_Q\,\frac{\dot{\overbar{R}}_ϕ(q^2)}{A_ϕ}\,\left(\frac{U''+ρ\,U^{(3)}}{\text{det}_{\text{B}}^2(Q)}
	+\frac{2\,ρ\,U''\left[ρ\,U''\left(U''+ρ\,U^{(3)}\right)-\left(2\,U''+ρ\,U^{(3)}\right)\,L_ϕ^S(Q)\right]}{\text{det}_{\text{B}}^3(Q)}\right)\,,\\[1ex]
	η_ϕ^{(B)} =&\, 4\,ρ\,\left(U''\right)^2\,\int_Q\,\frac{\dot{\overbar{R}}_ϕ(q^2)}{A_ϕ}\,\left(\frac{1+2\,\rBone+4\,q^2\,x^2\,\rBtwo}{\text{det}_{\text{B}}^2(Q)}-\frac{2\,q^2\,x^2\,\left(1+2\,\rBone\right)^2\,L_ϕ^S(Q)}{\text{det}_{\text{B}}^3(Q)}\right)\,.
\end{split}
\end{align}
\end{widetext}
After performing the Matsubara sums and the momentum integrations the overall flow equations in our truncation can be cast into the form
\begin{align}
\begin{split}
	\dot{U}^{(F)}(ρ) &= - \frac{16\,v_d}{d}\,k^{d+2}\,\ell_F^{(1,1)}\,\\[1ex]
	\dot{U}^{(B)}(ρ) &= \frac{8\,v_d\,2^{d/2}}{d} \,k^{d+2}\,\ell_B^{(1,1)}\,.
\end{split}
\end{align}
The fermionic contributions to the boson propagator are found to be
\begin{align}
\begin{split}
	\dot{S}_ϕ^{(F)} &= - \frac{16\,h^2\,v_d}{d}\,k^{d-4}\left(\ell_F^{(0,2)} -2\,w_3\,\ell_F^{(0,3)}\right)\,,\\[1ex]
	η_ϕ^{(F)} &=  \frac{16\,h^2\,v_d}{d}\,k^{d-4}\,\ell_{F,2}^{(0,2)}\,,
\end{split}
\end{align}
while the bosonic contributions are given by
\begin{align}
\begin{split}
	\dot{S}_ϕ^{(B)} =& -\frac{32\,S_ϕ}{d}\,ρ\,U''\,v_d\,2^{d/2}\,k^{d-4}
	\left[\vphantom{\int}\right.
	\left(U'' +ρ\,U^{(3)}\right)\,\ell_B^{(0,2)}\\
	&+2\,\left(ρ\,U''\right)^2\,\left(U''+ρ\,U^{(3)}\right)\,k^{-4}\,\ell_B^{(0,3)}\\[1ex]
  & -2\,ρ\,U''\,\left(2\,U'' +ρ\,U^{(3)}\right)\,k^{-2}\,\ell_B^{(1,3)}
	\left.\vphantom{\int}\right],\\[1ex]
	η_ϕ^{(B)} =& 8\,ρ\,\left(U''\right)^2\frac{v_d\,2^{d/2}}{d}\,k^{d-4}\,\ell_{B,2}^{(0,2)}\,.
\end{split}
\end{align}

Here we used the definitions for fermionic contributions
\begin{align}
	\ell_F^{(n,m)}\left(\tilde{μ},\tilde{T},w_3\right) =
	\begin{cases}
			\ell_2(\tilde{μ})\,\mathcal{F}_{\!\!m}\!\!\left(\sqrt{1+w_3}\right)\quad n\,\,\,\text{even}\\[1ex]
		\ell_1(\tilde{μ})\,\mathcal{F}_{\!\!m}\!\!\left(\sqrt{1+w_3}\right)\quad n\,\,\,\text{odd}
	\end{cases}
\end{align}
and
\begin{align}
	\ell_{F,2}^{(n,m)}\left(\tilde{μ},\tilde{T},w_3\right) =
	\begin{cases}
		\ell_3(\tilde{μ})\,\mathcal{F}_{\!\!m}\!\!\left(\sqrt{1+w_3}\right)\quad n\,\,\,\text{even}\\[1ex]
		\ell_1(\tilde{μ})\,\mathcal{F}_{\!\!m}\!\!\left(\sqrt{1+w_3}\right)\quad n\,\,\,\text{odd}
	\end{cases}\,,
\end{align}
where we made use of $w_3=h^2\,ρ/k^4$, as well as $w_1=U'/k^2$ and $w_2=ρ\,U''/k^2$. For bosonic diagrams we defined
\begin{widetext}
\begin{align}
	\ell_B^{(n,m)}\left(\tilde{T},w_1,w_2\right) = \frac{1}{S_ϕ^{2m}}\,\left(1-\frac{η_ϕ}{d+2}\right)\,\left(1+w_1+w_2\right)^n\,\mathcal{B}_{\!\!m}\!\!\left(\sqrt{(1+w_1)(1+w_1+2\,w_2)}/S_ϕ\right)
\end{align}
and
\begin{align}
	\ell_{B,2}^{(0,m)}\left(\tilde{T},w_1,w_2\right) =  \frac{1}{S_ϕ^{2m}}\,\mathcal{B}_{\!\!m}\!\!\left(\sqrt{(1+w_1)(1+w_1+2\,w_2)}/S_ϕ\right) = \left.\ell_B^{(0,m)}\right\rvert_{η_ϕ=0}\,.
\end{align}
\end{widetext}
$\mathcal{F}_{\!\!m}(z)$ and $\mathcal{B}_{\!\!m}(z)$ label the fermionic and bosonic Matsubara sums of order $m$, respectively. The functions $\ell_i$ are defined as
\begin{align}\label{eq:mom_F}
\begin{split}
	\ell_1(x) &= θ(x+1)\,(x+1)^{d/2} - θ(x-1)\,(x-1)^{d/2}\,,\\[1ex]
	\ell_3(x) &= θ(x+1)\,(x+1)^{d/2} + θ(x-1)\,(x-1)^{d/2}\,,\\[1ex]
  \ell_2(x) &= \ell_3(x)-2\,θ(x)\,x^{d/2}\,
	\end{split}
\end{align}
and the d-dimensional volume integral is given by $v_d^{-1} = 2^{d+1}\,π^{d/2}\,Γ(d/2)$.

In oder to obtain the flow of the density $n=n_{k\rightarrow0}$ we may split the chemical potential into a reference part $μ_0$ and an offset $δμ$ such that $μ=μ_0+δμ$. We then expand our effective potential \eqref{eq:effpot} with respect to  the offset chemical potential $δμ$ according to
\begin{align}
	U_k(ρ)&= \sum_{n = 1}^2\frac{u_n}{n!}\left(ρ-ρ_0\right)^n -n_k\,δμ+α_k\,(ρ-ρ_0)\,δμ
\end{align}
The differentiation with respect to $μ$ acts rather on $δμ$ as the reference chemical potential is fixed
\begin{align}
	\dot{n}_k=-\frac{\partial\,\dot{U}}{\partial\,δμ}\,.
\end{align}
According to our master equation for the effective average potential \eqref{eq:master_effpot} we now expand $L_ψ^{S,Q}$ and $\detF$ in terms of $δμ$ while the fermionic cutoff still regularises around the Fermi surface, i.e. the reference chemical potential $μ_0$.

\subsection{Flow equations for finite volume}
When confining our system by means of a compactifation of one spatial dimension in a dimensional crossover from 3d to 2d with a confinement length scale $L$.
By adopting periodic boundary conditions we restrict our system to a torus in one spatial direction
\begin{align}
	ψ(L) = ψ(0)
\end{align}
such that we obtain a ‘spatial Matsubara sum’ over discrete momenta $k_n = 2πn/L$ with $n\in\mathbb{Z}$.
Accompanying this quantisation of energy levels the bosonic and fermionic regulators defined are modified accordingly. For the optimised regulator they become
\begin{align}\label{eq:opt_reg_l}
\begin{split}
	&R_{ϕ,k}(q^2)= \left(k^2-\frac{q^2+k_n^2}{2}\right)\,θ\left(k^2-\frac{q^2+k_n^2}{2}\right)\,,\\
	&R_{ψ,k}(q^2)= k^2\,\left[\sgn\left(z+\tilde{k}_n^2\right) -\left(z+\tilde{k}_n^2\right)\right]\\
  &\hspace{2.5cm}\times\,θ\left(1-|z+\tilde{k}_n^2|\right)\,,
\end{split}
\end{align}
where we again used $z=(q^2-μ)/k^2$ and $\tilde{k}_n=k_n/k$.
Hence the $d$-dimensional spatial integration splits up into a sum over the discrete momenta $k_n$ and a momentum integral in $d-1$ dimensions
\begin{align}
	\int\,\frac{d^{d}\,q}{\left(2\,π\right)^{d}} = \frac{1}{L}\,\sum_{k_n}\,\int\,\frac{d^{d-1}\,q}{\left(2\,π\right)^{d-1}}\,.
\end{align}

Due to the inclusion of the discrete momenta in the regulator the evaluation of the spatial boils down to counting the modes within the potential well.
For periodic boundary conditions we hereby encounter the following type of sums
\begin{align}
	\begin{split}
		&\sum_{n=-N}^N α = α\, (1+2\,N)\qquad (α \in ℝ)\,,\\[1ex]
		&\sum_{n=1}^N n^2 = \frac{1}{6}\,N\,(1+N)\,(1+2\,N)\,,\\[1ex]\,
	\end{split}
\end{align}
and
\begin{align}
  \sum_{n=1}^N n^4 = \frac{1}{30}\,N\,(1+N)\,(1+2\,N)\,(-1+3\,N+3\,N^2)\,.
\end{align}
As a result of the periodic boundary conditions the regulator function restricts the Matsubara-type summation in the transversal direction to $|k_n| = |2πn/L|<\sqrt{2}\,k$ or equivalently $|n| < \tilde{L}/\sqrt{2}π$.

For bosonic contributions we define
\begin{align}
	N^{(B)} = \left\lfloor\frac{\tilde{L}}{\sqrt{2}\,π}\right\rfloor
\end{align}
with $\lfloor x \rfloor$ being the largest integer smaller than $x$.
In three dimensions we find
\begin{widetext}
\begin{align}
	\begin{split}
		C_L=&\frac{1}{L}\,\sum_{k_n}\,\left(1-\frac{k_n^2}{2\,k^2}\right)^{d/2}
	\left(1-\frac{η_ϕ}{d+2}\left(1-\frac{k_n^2}{2\,k^2}\right)\right)\,θ\left(k^2-\frac{k_n^2}{2}\right)	\\[1ex]
	=& \frac{k}{\tilde{L}}\left(1+2\,N^{(B)}\right)\,\left[
	1-\frac{η_ϕ}{4}-\frac{1}{6}\,\left(1-\frac{ηϕ}{2}\right)\,\left(\frac{2\,π}{\tilde{L}}\right)^2\,N^{(B)}\,\left(1+N^{(B)}\right)\right.\\
	&\hspace{1cm}\left.-\frac{η_ϕ}{60}\,\left(\frac{2\,π}{\tilde{L}}\right)^4\,N^{(B)}\,\left(1+N^{(B)}\right)\,\left(-1+3\,N^{(B)}+3\,\left(N^{(B)}\right)^2\right)\right].
	\end{split}
\end{align}
\end{widetext}
Thus all bosonic flow equations still hold with the replacements
\begin{align}
	\left(1-\frac{η_ϕ}{d+2}\right)\rightarrow C_L \,,\qquad d\rightarrow d-1.
\end{align}

The fermionic momentum integrals can be generalised by the transformation $z\rightarrow \widehat{z}=(q^2+k_n^2-μ)/k^2$. All results can then be transferred by the transformation $μ\rightarrow\widehat{μ} = \tilde{μ}-\tilde{k_n^2}$.
For periodic boundary conditions it can be easily shown in $d=3$ dimensions
\begin{widetext}
\begin{align}
\begin{split}
	&\frac{1}{L}\sum_{k_n}\,θ\left(\hat{μ}+1\right)\,\left(\hat{μ}+1\right)^{(d-1)/2}= \frac{1}{L} \left[ \left(\tilde{μ} +1\right)\,\left(1+2\,N_1^{(F)}\right)-\frac{1}{3}\,\left(\frac{2\,π}{\tilde{L}}\right)^2\,\nfo\,\left(1+\nfo\right)\,\left(1+2\,\nfo\right)\right]\,θ\left(\tilde{μ}+1\right)\,,\\[1ex]
	&\frac{1}{L}\sum_{k_n}\,θ\left(\hat{μ}-1\right)\,\left(\hat{μ}-1\right)^{(d-1)/2}= \frac{1}{L} \left[ \left(\tilde{μ} -1\right)\,\left(1+2\,N_2^{(F)}\right)-\frac{1}{3}\,\left(\frac{2\,π}{\tilde{L}}\right)^2\,\nft\,\left(1+\nft\right)\,\left(1+2\,\nft\right)\right]\,θ\left(\tilde{μ}-1\right)\,,\\[1ex]
		&\frac{1}{L}\sum_{k_n}\,θ\left(\hat{μ}\right)\,\left(\hat{μ}\right)^{(d-1)/2}=\frac{1}{L} \left[ \tilde{μ}\,\left(1+2\,N_3^{(F)}\right)-\frac{1}{3}\,\left(\frac{2\,π}{\tilde{L}}\right)^2\,\nftt\,\left(1+\nftt\right)\,\left(1+2\,\nftt\right)\right]\,θ\left(\tilde{μ}\right).
	\end{split}
\end{align}
\end{widetext}
Here we defined
\begin{align}
  \begin{split}
	N_1^{(F)} &= \left\lfloor\frac{\tilde{L}\,\left(\tilde{μ}+1\right)^{1/2}}{2\,π}\right\rfloor\,,\\[1ex]
	N_2^{(F)} &= \left\lfloor\frac{\tilde{L}\,\left(\tilde{μ}-1\right)^{1/2}}{2\,π}\right\rfloor\,,\\[1ex]
	N_3^{(F)} &= \left\lfloor\frac{\tilde{L}\,\tilde{μ}^{1/2}}{2\,π}\right\rfloor\,.
\end{split}
\end{align}
Hence for the spatial threshold function with explicit Matsubara summation we obtain for periodic boundary conditions in $d=3$
\begin{widetext}
\begin{align}
	\begin{split}
		\frac{1}{L}\sum_{k_n}\,\ell_a(\hat{μ})
		= &\frac{k}{\tilde{L}}\left\{
		\left[ \left(\tilde{μ} +1\right)\,\left(1+2\,N_1^{(F)}\right)-\frac{1}{3}\,\left(\frac{2\,π}{\tilde{L}}\right)^2\,\nfo\,\left(1+\nfo\right)\,\left(1+2\,\nfo\right)\right]\right.\\[1ex]
		&(-1)^a \left[ \left(\tilde{μ}+1\right)\rightarrow\left(\tilde{μ}-1\right) \& \left(\nfo\rightarrow\nft\right)\vphantom{\frac{k}{\tilde{L}}}\right]\\[1ex]
		&-(1+(-1)^a)\left.\left[ \tilde{μ}\,\left(1+2\,N_3^{(F)}\right)-\frac{1}{3}\,\left(\frac{2\,π}{\tilde{L}}\right)^2\,\nftt\,\left(1+\nftt\right)\,\left(1+2\,\nftt\right)\right]\,\right\}
	\end{split}
\end{align}
for $a=1,2$ and in addition
\begin{align}
	\begin{split}
		\frac{1}{L}\sum_{k_n}\,\ell_3(\hat{μ})
		= &\frac{k}{\tilde{L}}\left\{
		\left[ \left(\tilde{μ} +1\right)\,\left(1+2\,N_1^{(F)}\right)-\frac{1}{3}\,\left(\frac{2\,π}{\tilde{L}}\right)^2\,\nfo\,\left(1+\nfo\right)\,\left(1+2\,\nfo\right)\right]\right.\\[1ex]
		&+ \left.\left[ \left(\tilde{μ}+1\right)\rightarrow\left(\tilde{μ}-1\right) \& \left(\nfo\rightarrow\nft\right)\vphantom{\frac{k}{\tilde{L}}}\right]\right\}\,
	\end{split}
\end{align}
\end{widetext}
Thus all fermionic flow equations can be transferred to the case of finite volume with periodic boundary conditions with the replacement
\begin{align}
	\ell_i \rightarrow \ell_{i,L} = \frac{k}{\tilde{L}}\,\sum_{k_n}\,\ell_i\,,\qquad d\rightarrow d-1.
\end{align}

\section{Numerical procedure}\label{ap:numerics}

The set of coupled differential equations for the projected flow equations from Appendix \ref{ap:floweq} are numerically evaluated for both zero and finite temperature. However, it is a useful feature of the functional renormalisation group that for larges scales $k^2\gg T$ the finite temperature flow can be approximated by the zero temperature system \cite{PhysRevA.89.053630}. For a practical computation we choose $k_{\text{switch},T}=6\,π\,T$, i.e. we follow the zero temperature flow until $k_{\text{switch},T}$ where the temperature starts to become an important scale and we switch to the finite temperature flow equations.

Likewise, the Fermi gas confined to a trap can be regarded as an unconfined system for large scales $k\gg L^{-1}$. Here we choose $k_{\text{switch},L}=50/L$, which significantly decreases the runtime of the computation. The agreement of the results with and without splitting the flow in zero and finite temperature, as well as unconfined and confined flow equations was checked numerically.

The critical temperature is determined as the largest temperature for which the gap of the fermion spectrum is non-vanishing. Numerically, we use the following algorithm
\begin{align}
  0 < Δ_{t_{\text{final}}}\left(T_c,μ,a,L\right) < \frac{1}{100}Δ_{t_{\text{final}}}\left(T=0,μ,a,L\right)\,.
\end{align}
Using this algorithm is very effecient as it accounts for both the large gap on the BEC-side, as well for the smaller gap on the BCS-side (especially in the three-dimensional case). It was checked numerically that further limitation to $<1\%$ of the zero temperature gap yields identical results in $d\leq 3$ within the numerical precision.

\bibliographystyle{apsrev4-1}
\bibliography{Bibliography_DimXover}

\end{document}